\documentclass[12pt]{article}
\usepackage[utf8]{inputenc}
\usepackage{amsmath, amsthm, amssymb, fancyhdr, fancybox, graphicx}
\usepackage{multirow}
\usepackage{pdflscape}
\usepackage{rotating}
\usepackage{booktabs}
\usepackage[authoryear]{natbib}
\usepackage[english]{babel}
\usepackage{floatrow}
\usepackage{caption}
\usepackage{authblk}
\usepackage{bbm}

\begin{document}
\title{\Large \bf A Copula-based Imputation Model for Missing Data of Mixed Type in Multilevel Data Sets}
\author[1]{\small Jiali Wang\thanks{Corresponding author: Jiali Wang, Research School of Finance, Actuarial Studies and Statistics, College of Business and Economics Building 26C, The Australian National University, Canberra, ACT 2601, Australia. Phone: +61 2 612 57290. Email: u5298171@anu.edu.au}}
\author[1]{\small Bronwyn Loong}
\author[1,2]{\small Anton H. Westveld}
\author[3]{\small Alan H. Welsh}

\affil[1]{\scriptsize Research School of Finance, Actuarial Studies and Statistics, College of Business and Economics, The Australian national University, AUS} 
\affil[2]{\scriptsize Statistics Laboratory $@$ the Bio5 Institute \& Statistics G.I.D.P., The University of Arizona, USA}
\affil[3]{\scriptsize Mathematical Sciences Institute, College of Physical \& Mathematical Sciences, The Australian National University, AUS}
\date{}
\maketitle
\fancyfoot[L]{\today\ \currenttime}

\begin{abstract}
We propose a copula based method to handle missing values in multivariate data of mixed types in multilevel data sets. Building upon the extended rank likelihood of \cite{hoff2007extending} and the multinomial probit model, our model is a latent variable model which is able to capture the relationship among variables of different types as well as accounting for the clustering structure. We fit the model by approximating the posterior distribution of the parameters and the missing values through a Gibbs sampling scheme. We use the multiple imputation procedure to incorporate the uncertainty due to missing values in the analysis of the data. Our proposed method is evaluated through simulations to compare it with several conventional methods of handling missing data. We also apply our method to a data set from a cluster randomized controlled trial of a multidisciplinary intervention in acute stroke units. We conclude that our proposed copula based imputation model for mixed type variables achieves reasonably good imputation accuracy and recovery of parameters in some models of interest, and that adding random effects enhances performance when the clustering effect is strong.
\end{abstract}

\section{Introduction}
Multivariate analysis often involves understanding the relationship among variables of different types. Our motivating data set is from the Quality in Acute Stroke Care (QASC) study, which implemented a multidisciplinary intervention to manage fever, hyperglycaemia and swallowing dysfunction in acute stroke patients \citep{middleton2011implementation}. This study was one of the largest rigorously evaluated clinical trials which showed that organised stroke unit care significantly reduced death and disability among stroke patients. There were 19 acute stroke units in New South Wales, Australia that participated in the study, and they were randomly assigned to an intervention group (10 units) and a control group (9 units). A pre-intervention and a post-intervention cohort of patients were recruited , their demographic variables such as age, gender and marital status were obtained, and process of care variables such as temperature, time from onset to hospital and length of stay in hospital were recorded. The researchers were interested to see if the implementation of the protocols reduced death and dependency, and improved physical and mental health scores. The four primary outcome variables considered were: (1) modified Rankin Scale (an ordinal variable ranging from 0 to 6, measuring the degree of disability or dependence in daily activities); (2) Barthel index (an ordinal variable ranging from 0 to 100, which also measures performance in activities of daily living. It is usually reported as a dichotomised variable with 60 or more and 95 or more as cut points); (3) mean SF-36 mental component summary score; (4) mean SF-36 physical component summary score. Mental and physical component summary scores were measured on continuous scales between 0 and 100.

In the QASC study, all the four outcome variables had moderate amounts of missing data and most of the explanatory variables had missing values as well (Table 1). Ignoring all the patients with missing values, which is known as complete case analysis, is a commonly used approach to handle missing data but may lead to biased estimates and reduced statistical power. In other words, the remaining cases may not be representative of the target population if we ignore them completely. The smaller sample size also decreases the power to detect significant treatment effects. Due to the potential for positive dependence among units within the same cluster, this is especially serious in multilevel data sets. Case-wise deletion reduces the sample size of patients within hospitals and the number of hospitals at the same time if any information at the hospital level is missing. As a consequence, both the variations between and within hospitals may not be accurately estimated. An alternative approach  is to `impute' missing values, so that after imputation complete data analysis can be performed using standard software. Some ad-hoc procedures include mean imputation and last observation carried forward.  More principled imputation methods are model-based, such as joint modelling \citep[chapter 11]{little2014statistical} and fully conditional specification \citep{raghunathan2001multivariate,van2007multiple}.

\begin{table}
\scalebox{0.8}{
\begin{tabular}{ |c|c|c|c|c|}
\hline
Variable group &Variable Names & Variable Type &Missing Percentage \\
\hline\hline
\multirow{4}{6em}{Outcomes} & modified Rankin Scale &ordinal&9.48\%\\
&Bartell Index& ordinal &15.14\%\\
& physical health score& continuous &15.74\% \\
&mental health score& continuous &15.74\%\\
\hline
\multirow{4}{6em}{Allocations} & hospcode &indicator&0\%\\
&id& indicator &0\%\\
&treatment& binary &0\% \\
&period& binary &0\%\\
\hline
\multirow{5}{6em}{Demographic} & gender &binary&0\%\\
&age &continuous &5.89\%\\
&marital Status& nominal &14.8\%\\
&highest education level& ordinal &15.95\% \\
&ATSI& binary &17\%\\
\hline
\multirow{3}{6em}{Process of Care} & time to presentation  &continuous&1.69\%\\
&length of stay &count &4.53\%\\
&mean temperature& continous &4.73\%\\
\hline
\end{tabular}}
\caption {Summary of variables in the QASC}
\end{table}

Current methods to handle missing data are potentially inadequate to apply to the QASC study which is complicated by the clustering effect and the mix of variable types. \cite{hoff2007extending} proposed using a semiparametric copula model based on the extended rank likelihood to analyse multivariate data of mixed types. We extend the work of \cite{hoff2007extending} by adding random effects to introduce correlation among individuals within clusters. The model in \citep{hoff2007extending} can only be used for continuous and ordinal variables, so we consider a multinomial probit model to handle nominal variables. We then evaluate our model by its ability to recover missing data and estimate the true parameters in some models of interest in both a simulation study and a real data study.

The structure of this manuscript is as follows. In section 2 we briefly summarize some popular multivariate techniques to perform missing data imputation and review the general Gaussian copula model and the extended rank likelihood for semiparametric copula estimation as discussed in \cite{hoff2007extending}. In section 3 we describe this extended rank likelihood with random effects and combine the copula model with a multinomial probit model. We outline our algorithm to impute missing data in a multilevel data set using our proposed copula model. In section 4, we present and discuss the results of our simulation and real data studies to evaluate our model. The proposed model is compared against several conventional methods using readily available software packages.  Section 5 provides concluding remarks and discusses some future research.

\section{Background of Missing Data Imputation}
Let $Y=(Y_{obs},Y_{mis})$ denote the `complete' data, with observed part $Y_{obs}$ and missing part $Y_{mis}$. Let $\theta$ denote the parameter describing the `complete' data $Y$. Throughout this paper we assume the data are Missing at Random (MAR)\citep{rubin1976inference}, meaning that the probability of missing an entry only depends on the observed data, not on the entry value itself, so that inference about $(Y_{mis},\theta)$ can be made based on only the observed data $Y_{obs}$, and no extra effort is needed to model the missing data process \cite[p.12]{schafer1997analysis}. The MAR assumption cannot be tested except in artificial simulation settings, however, it is a simplifying assumption which can be made more reasonable by expanding the model to include more variables that are related to the missing data. Data augmentation \citep{tanner1987calculation} is often used as a simulation based computational algorithm to approximate the joint posterior distribution of $p(\theta,Y_{mis}|Y_{obs})$. It draws $Y_{mis}$ from $p(Y_{mis}|Y_{obs},\theta)$ and $\theta$ from $p(\theta|Y)$ iteratively. The $\theta$ can be treated as coming from the marginal distribution $p(\theta|Y_{obs})$ and the $Y_{mis}$ can be treated as coming from $p(Y_{mis}|Y_{obs})$, if our interest lies in filling in the missing values to create complete data sets.

\subsection{Multiple Imputation}
Having obtained guesses for the missing data from an imputation model (which will be discussed further below), we cannot treat them as the `true' data because of the uncertainty due to nonresponse. \cite{rubin1987multiple} proposed multiple imputation (MI) to obtain $M$ independent draws of $Y_{mis}$ from $p(Y_{mis}|Y_{obs})$ to create $M$ complete data sets. Combining rules are then applied to the parameter estimates from each of the $M$ complete data sets to obtain a single inferential result, as follows.

Let $Q$ be the target population quantity of interest, for example, the coefficients of a regression model. Suppose $\hat q_m$ is the point estimate of $Q$ from the $m^{th}$ imputed complete data set and $\hat w_m$ is an associated measure of sampling variance, $m=1,...,M$. Three quantities are required for inference on $Q$:
\begin{equation}\label{combining rule}
\begin{aligned}
\bar q&=\frac{1}{M} \sum_{m=1}^{M} \hat q_m\\
B&=\frac{1}{M-1} \sum_{m=1}^{M} (\hat q_m-\bar q)^2\\
\bar W&=\frac{1}{M} \sum_{m=1}^{M} \hat w_m
\end{aligned}
\end{equation}
The analyst uses $\bar q$ as the point estimate of $Q$. The sampling variance of $\bar q$ is estimated by $T=\bar W+ (1+ \frac{1}{M}) B$. The total variance associated with $\bar q$ is a function of the within imputation variance and the between imputation variance.

Next we discuss common approaches to impute missing values.

\subsection{Approaches to Generate Imputations for Missing Values}
A good imputation method aims to preserve relationships among survey variables of interest. The joint modelling (JM) approach usually assumes the data follow an elliptical joint distribution, for example, a multivariate normal or a multivariate $t$ distribution. For continuous variables, some transformations may be needed to approximate the assumed distribution \citep{goldstein2009multilevel}. Discrete variables are treated as if they were generated from the underlying continuous variables and then discretized. Most software packages implement the joint modelling approach by first transforming any variables with missing values into responses that follow a multivariate normal distribution. The transformed responses are then regressed against the fully observed variables. The software packages that implement this approach include \textit{norm} \citep{fox2013package} and \textit{Amelia} \citep{honaker2011amelia} in R and PROC MI in SAS. Other joint modelling techniques include loglinear models and general location models specifically designed for categorical data and mixed data respectively \citep{little2014statistical}. Another useful package in R - \textit{pan} \citep{schafer2002computational} is designed to impute missing values in panel data, assuming a multivariate Gaussian distribution with random effects. \cite{goldstein2009multilevel} further extended Schafer's multilevel imputation model by allowing for multivariate response variables at all levels of a data hierarchy, and used Box-Cox type normalizing transformations for continuous non-Gaussian responses. Although elliptical distributions allow for parsimonious description of data, they are restrictive in the marginal distributions which are fully determined by the parent joint distribution and are restrictive in capturing complex dependencies among variables.

The fully conditional specification (FCS) \citep{raghunathan2001multivariate,van2007multiple} approach breaks the joint model into a series of univariate regression models. Generalized linear models are often specified to accommodate different types and shapes of variables as well as adding constraints among variables. This method has been implemented by many software packages, for instance, \textit{mice} \citep{buuren2011mice} and \textit{mi} \citep{su2011multiple} in R, \textit{ice} in STATA \citep{royston2005multiple} and a SAS-based software IVEware \citep{raghunathan2002iveware}. To the best of our knowledge, there are no available packages to implement the multilevel fully conditional specification except for the `mice.impute.2l.norm' function in the \textit{mice} package in R, which fits mixed effects linear regression models for variables with missing values. Because of the lack of packages for practitioners, some authors have investigated including indicator variables for clusters \citep{drechsler2015multiple,eddings2011accounting} in the imputation models or ignore the clustering effects. The main criticism of the fully conditional methods, however, is the lack of theoretical justification to ensure the univariate conditional distributions converge to a proper joint distribution.

Several papers have compared JM and FCS MI, but there is no clear conclusion under which circumstances practitioners should favour one over the other. \cite{lee2010multiple} performed simulations under three missing data mechanisms and their results showed that JM and FCS produce similar results despite the data not being multivariate normal. \cite{kropko2013multiple} not only assessed the accuracy of the coefficients fitted to models of interest, but also the accuracy of imputed values. Their study found that FCS imputed more accurately for categorical variables than JM but the differences were small for continuous variables. \cite{zhao2009performance} studied the performance of JM and FCS in multilevel settings, and showed using simulations that FCS MI outperforms JM MI in having less bias, and when the intraclass corrlation is small, more accurate parameter estimates are obtained from both JM and FCS.

\subsection{Copulas}
To provide more flexibility in the marginal distributions while at the same time ensuring a proper joint distribution, we consider copula modelling approaches to impute missing values. The word `copula' means `a link, tie, bond'. In mathematics and statistics, it means joining together one-dimensional distribution functions to form a multivariate distribution function. Specifically, the distribution functions for the random variables $y_1,...,y_p$ are $F_1(y_1),...,F_p(y_p)$. Sklar's theorem \citep{sklar1959fonctions} shows that there always exists a function $C$, such that, $F(y_1,...,y_p)=C(F_1(y_1),...,F_p(y_p))$, where the function $C$ is called the copula function. Each of the variables is modeled by the marginal distribution $F_l(y_l)=u_l$, $l=1,...,p$, which is uniformly distributed, and their dependence is captured by the copula function $C$. Copula modelling has proven to be very powerful for modeling variables of different types and shapes, when there is an underlying dependence among them. It adopts a `bottom-up' strategy where the starting point is the marginal distributions $F_l$, which are then glued together by the copula function $C$. In the `top down' joint modelling approach, the marginal distributions are fully determined by their parental joint distribution so that there is no flexibility to model them. In addition, copula models guarantee the existence of a compatible joint distribution which is not guaranteed by the fully conditional specification approach. Existing models, like multinomial (ordered) probit models for (ordered) categorical data can be treated as special cases of copulas, because the underlying latent variables corresponding to each category are assumed to follow a multivariate Gaussian distribution \citep{chib1998analysis}.

In a copula model, the parameters are the marginal distributions $F_l$ and the copula function $C$. \cite{pitt2006efficient} developed a fully Bayesian estimation procedure to model the joint distribution of both sources of parameters. However, specifying each of the marginal distributions is labour intensive and variables in real data sets may not be accurately represented without a large number of parameters. Some authors suggested transforming the variables using the empirical distribution $\hat F_l$ to get pseudo data \citep{genest1995semiparametric} and avoid the parametric estimation of marginal distributions. However, this only applies to continuous variables. To link the discrete variables with continuous latent variables, \cite{hoff2007extending} provided a simple way of analysing the correlation among variables with meaningful ordering (continuous and ordered categorical variables), via the extended rank likelihood. This makes use of the fact that the order of the underlying latent variable is consistent with the observed data, and inference about the association parameters can be drawn from the `rank-based' latent variables through a simple parametric form. The extended rank likelihood has already been applied to other closely related models, for example, a general Bayesian Gaussian copula factor model proposed by \cite{murray2013bayesian} and a bifactor model considered by \cite{gruhl2013semiparametric}, can be treated as imposing a special structure on the correlation matrix of a Gaussian copula.

Using the copula model as an imputation engine is relatively new but has drawn some attention in the literature. \cite{kaarik2009modeling} were among the first authors to consider imputation using a Gaussian copula where the missing data pattern was monotone. \cite{lascio2015exploring} found that copula based imputation from the Archimedian family compared favourably with nearest neighbour donor imputation and regression imputation by the EM algorithm. \cite{hollenbach2014fast} compared the performance of imputation by the copula model using the extended rank likelihood approach \citep{hoff2007extending} with JM (as implemented in Amelia) and FCS (as implemented in MICE) and concluded that the copula imputation approach maintained the prediction accuracy at least as well as the other two approaches but with faster convergence of the sampling algorithm.

\section{Semi-parametric Gaussian copula model}
\subsection {The Extended Rank Likelihood with Random Effects}
Among a variety of copulas, we focus on the Gaussian copula in this paper. For further theoretical details of copulas, see \cite{nelsen2007introduction} and for a good summary of some applications of copulas, see \cite{trivedi2007copula}.  Rather than assuming a Gaussian distribution on the data $Y$ directly, the Gaussian copula specifies a joint multivariate Gaussian distribution on the corresponding latent variables as defined next. Let $l=1,...,p$ denote the index of the $l^{th}$ random variable. Then the $l^{th}$ latent variable is $z_l=\Phi^{-1}(u_l)$, where $u_l=F_l (y_l)$. That is, $C(u_1,...,u_p|\Gamma)=\Phi_p(\Phi^{-1}(u_1),...,\Phi^{-1}(u_p)|\Gamma)=\Phi_p(z_1,...,z_p|\Gamma)$, where $\Phi_p(\cdot|\Gamma)$ is the cumulative distribution function of the p-variate normal distribution, with mean zero and correlation matrix $\Gamma$. Note that the Gaussian copula can reach the full range of pairwise correlation (-1,1) and the parameters that need to be estimated only come from the correlation matrix $\Gamma$.

\cite{hoff2007extending} derived a rank-based likelihood to estimate the correlation matrix $\Gamma$ so that there is no need to specify the marginal distributions $F_l$. The idea is that since we know $\Phi^{-1}(F(\cdot))$ is a monotone transformation, the ordering of data $Y$ provides partial information about what $z$ should be, that is, $y_{i_1l}<y_{i_2l}$ implies $z_{i_1l}<z_{i_2l}$. Suppose we have in total $N$ observations, $n=1,...,N$. Observing $y=(y_1,...,y_N)$ tells us that $z=(z_1,...,z_N)$ must lie in the set: $\big\{z \in \mathbb{R} ^{N \times p}: max\{z_{hl}: y_{hl}<y_{nl}\}< z_{nl} < min \{z_{hl}:y_{hl}> y_{nl}\} \big\}$. Let `$D$' denote the set of all possible $z$ which is consistent with the ordering of $y$. Then the event `$z \in D$' can be treated as the observed event upon which inference of $\Gamma$ is made. The full likelihood can be decomposed as
\begin{equation}\label{full likelihood}
 \begin{aligned}
  p(y|\Gamma, F_1,...,F_p)&=p(z\in D, y|\Gamma, F_1,...,F_p)\\
  &=p(z\in D|\Gamma) \times p(y|z\in D,\Gamma,F_1,...,F_p).
  \end{aligned}
\end{equation}

\cite{hoff2007extending} proved that it is partial sufficient (in the sense of G-sufficient and L-sufficient) to carry out inference about $\Gamma$ based on the density $p(Z\in D|\Gamma)$ and he referred to it as the `extended rank likelihood'. In doing so, we lose the information about $\Gamma$ from the density $p(y|z\in D,\Gamma,F_1,...,F_p)$, but we do not need to estimate the potentially complicated marginal distribution functions and the extended rank likelihood provides a more general and flexible framework for joint modelling.

To take into account clustering effects, we extend Hoff's work by adding random effects to the Gaussian copula model at the latent variable level. The idea is that the clustering of the observed data is carried through to the latent variable level.  Our model can be described as
\begin{equation}\label{extended copula}
  z_{ij}|b_{i1} \sim N_p(b_{i1},\Gamma_1), b_{i1} \sim N_p(0,\Psi_1),
\end{equation}
where $i=\{1,...,m\}$ is the group index, $j=\{1,...,n_i\}$ is the individual index within group $i$, $\Gamma_1$ is a correlation matrix and $\Psi_1$ is a variance-covariance matrix for $z_{ij}$ and $b_{i1}$. Both $z_{ij}$ and $b_{i1}$ are vectors of length $p$, because we are considering $l=\{1,...,p\}$ variables jointly. In this model, the parameters that need to be estimated are in $(\Gamma_1,\Psi_1)$, which can be thought of as splitting the total correlation into two parts, the variability within groups and the variability between groups. However, like any model that relies on the ordering of the data but not their magnitude,  model (\ref{extended copula}) suffers from an identifiability problem without constraints on $\Gamma_1$. To see this, if we shift the location of the latent variable $z_l$ by $\mu_l$ and scale it by $\sigma_l$, the model remains unchanged because the new latent variables satisfy the order of the observed data as well. The extended rank likelihood contains only the information about the relative ordering of $z$ but no information about their location and scale. To solve the identifiability problem of scale, we fix $\Gamma_1$ to be a correlation matrix instead of a covariance matrix. In other words, there is no need to estimate the variances of $z$ conditional on the random effects, so we fix them as 1. Because the marginal distributions of $z$ have mean equal to 0, there is no identifiability issue for location. We will briefly describe how to add covariates in the discussion section so that the mean of $z$ is no longer 0.

\subsection{Copula Model for Mixed Type Variables}
Notice that the extended rank likelihood described above only applies to continuous and ordinal variables, since it makes no sense to consider meaningful numeric values for nominal variables (categorical variables without ordering). To include nominal variables in the copula model as well, we consider a multinomial probit model \citep{aitchison1970polychotomous, chib1998analysis} which can be treated as a Gaussian copula. The idea is to relate a nominal variable to a vector of latent variables which can be thought of as the unnormalized probabilities of choosing each of the categories. Suppose a single nominal variable $y$ has $K$ categories, and we define $K-1$ latent variables for unit $i$ as $w_i=(w_{i1},...,w_{i,K-1})$ which follow a multivariate Gaussian distribution. Since all the variables appear on one side and we model them jointly, there are no covariates as predictors for now, therefore we only include the intercept term $\beta$ vector to represent the relative differences between each category $1,...,K-1$ compared with the baseline category $K$. To add a second level to the hierarchy, again we have the random effects $b_{i2}$ in the model
\begin{equation}\label{multinomial probit model}
\begin{aligned}
  w_{ij}=&\beta+b_{i2}+\epsilon_{ij}\\
  b_{i2} \sim & N_{K-1}(0,\Psi_2), ~\epsilon_{ij} \sim N_{K-1}(0,\Gamma_2)\\
  y_{ij} =&\begin{cases}k~~ \text{if}~ w_{ijk} >w_{ijk'}~\text{and}~ w_{ijk}>0,~\text{for}~k'\neq k \\
      K ~\text{if}~ w_{ijk}<0, ~\text{for all}~k=1,...,K-1. \end{cases}
\end{aligned}
\end{equation}
The rule of deciding the category is a mapping from the latent variables vector to the observed category. The category $k=1,...,K-1$ is observed if the $k^{th}$ element of the vector $w_i$ is the largest and greater than 0; the last category $K$ is observed if the largest element in $w_i$ is smaller than 0. We also fix the diagonal elements of $\Gamma_2$ equal 1 to be identifiable.

To provide a unified framework of multivariate analysis for mixed type variables, we combine model (\ref{extended copula}) for variables with ordering and model (\ref{multinomial probit model}) for variables without ordering as follows

\begin{equation}\label{combined}
\begin{aligned}
  &z_{ij}|b_{i1}\sim N_p(b_{i1},\Gamma_1), w_{ij}\sim N_{K-1}(\beta+b_{i2},\Gamma_2),\\
  &b_i=(b_{i1},b_{i2})\sim N_{p+K-1}(0,\Psi), \Psi=
   \begin{pmatrix}
  \Psi_1 & \Psi_{12} \\
  \Psi_{21} &  \Psi_2
 \end{pmatrix},\\
  &(z_{ij},w_{ij})|b_i\sim N_{p+K-1}((0,\beta)+b_i,\Gamma), \Gamma=
  \begin{pmatrix}
  \Gamma_1 & \Gamma_{12} \\
  \Gamma_{21} &  \Gamma_2
 \end{pmatrix}.
\end{aligned}
\end{equation}
The correlations between variables $y_1,...,y_p$ and $y_{p+1}$ are modelled through the off-diagonal matrices $\Psi_{12}$ and $\Gamma_{12}$ at the group level and the individual level respectively. Since both $\Gamma_1$ and $\Gamma_2$ have diagonal elements fixed to be 1, the big matrix $\Gamma$ is an identifiable correlation matrix.

\subsection{A Gibbs Sampler}
 A Gibbs sampling scheme is constructed to approximate the joint posterior distribution $p(\beta,\Psi,\Gamma,b,z,w,y_{mis}|y_{obs})$ where the unknown quantities in model (\ref{combined}) are the parameters $(\beta,\Psi,\Gamma)$ and the latent variables $(b,z,w)$ as well as missing data $y_{mis}$. A simple conjugate prior does not exist for a correlation matrix, and we follow the idea in \cite{hoff2007extending} of employing a parameter expansion approach \citep{liu1999parameter} to facilitate the MCMC sampling. Specifically, we put an Inverse Wishart prior on the matrix $\tilde \Gamma$ which is the semi-conjugate prior in a multivariate Gaussian sampling model. Then the full conditional distribution of $\tilde \Gamma$ can be derived analytically. After updating $\tilde \Gamma$ in each iteration, we rescale it to be a correlation matrix $\Gamma$. For ease of computation, we put an improper flat prior on $\beta$ and a semi-conjugate Inverse Wishart prior on $\Psi$, where the hyperparameters are the degrees of freedom $\nu$ and the scale matrix $\Lambda$
\begin{equation}\label{priors}
\begin{aligned}
&p(\beta)\propto 1, \\
&\Psi \sim Inv~Wishart (\nu_1,\Lambda_1),\\
&\tilde \Gamma \sim Inv~Wishart (\nu_2,\Lambda_2).
\end{aligned}
\end{equation}

Under these priors, it is straightforward to derive the full conditional distributions for the parameters $(\beta,\Gamma,\Psi)$ as follows
\begin{enumerate}
  \item $p(\beta|\dots) \sim N(\frac{1}{N} \sum_{i=1}^m \sum_{j=1}^{n_i}(w_{ij}-b_{2i}-\Gamma_{21}\Gamma_1^{-1}(z_{ij}-b_{i1})),\frac{1}{N}(\Gamma_2-\Gamma_{21}\Gamma_1^{-1}\Gamma_{12}))$;
  \item $p(\tilde \Gamma|\dots) \sim Inv~Wishart (\nu_1-1+N,\Lambda_1+\sum_{i=1}^m\epsilon_i^T\epsilon_i)$, where $\epsilon_i=(z_i,w_i)-(0,\beta)-b_i$\\
  $\Gamma_{[g,h]}=\tilde \Gamma_{[g,h]}/\sqrt{\tilde \Gamma_{[g,g]}\tilde \Gamma_{[h,h]}}, g,h=1,...,p$, $\Gamma$ is rescaled from $\tilde \Gamma$ after each sampling;
  \item $p(\Psi|\dots)\sim Inv~Wishart (\nu_2+m,\Lambda_2+B^T B)$, where $B=(b_1,...,b_m)$.

From the joint Gaussian distribution of $(z,w)$, we can derive the following conditional distributions for the latent variables $z,w,b$:

  \item $p(z_{ij}|\dots) \sim N(b_{1i}+\Gamma_{12}\Gamma_2^{-1}(w_{ij}-\beta-b_{i2}),\Gamma_1-\Gamma_{12}\Gamma_2^{-1}\Gamma_{21})$;
  \item $p(w_{ij}|\dots) \sim N(\beta+b_{2i}+\Gamma_{21}\Gamma_1^{-1}(z_{ij}-b_{i1}),\Gamma_2-\Gamma_{21}\Gamma_1^{-1}\Gamma_{12})$;\\
  $z_{ij}$ and $w_{ij}$ should be sampled from a truncated Gaussian distribution and a Gaussian distribution under the observed category constraint respectively, see below for details.
  \item $p(b_i|\dots) \sim N (U_i(\Gamma^{-1}\otimes 1_{n_i}^T)vec((z_i,w_i)-(0,\beta)),U_i)$,where $U_i=(\Psi^{-1}+n_i\Gamma^{-1})^{-1}$.
\end{enumerate}

The operator $\otimes$ is the Kronecker product and $vec()$ is the operator that vectorizes a matrix by stacking its columns. Updating the latent variable $z$ is achieved by sampling from a truncated multivariate Gaussian distribution, where the lower and upper bounds for each single entry $z_{ijl}$ are determined by: $lw=max(z_{hl}:y_{hl}<y_{ijl})$ and $up=min(z_{hj}:y_{hj}>y_{ijl})$ respectively, and $h$ is the index that searches over all the rows in the $l^{th}$ variable. For example, the lower bound for $z_{ijl}$ is the maximum value of the latent variable $z$ in the $l^{th}$ column whose corresponding $y$ is smaller than $y_{ijl}$ and the upper bound can be defined accordingly. Updating the latent variable $w$ is achieved by sampling from a multivariate Gaussian distribution under the constraint of the observed category by an acceptance and rejection algorithm \citep{albert1993bayesian}. Specifically, we sample a $w$ vector from the multivariate Gaussian distribution and accept this draw if and only if the maximum element of $w$ occurs at the place of the observed category and is greater than 0, or all the elements in $w$ are smaller than 0 and we observe the reference category $K$. We continue to sample $w$ until a draw is accepted. When there are missing values in $(y_1,...,y_p)$, the lower and upper bounds in $z$ are undefined, and/or any missing value occurs in $y_{p+1}$, the observed category in $y_{p+1}$ does not exist. In these cases, we just sample $z$ and/or $w$ from the multivariate Gaussian distributions without the constraints.

To sample missing values for variables with ordering, we apply the monotone transformation on $z$: $y_{ijl}=\hat F_l^{-1}[\Phi(z_{ijl})], l=1,...,p$, where $\hat F_l$ is the univariate empirical distribution function of variable $y_l$. To sample the missing values in nominal variables, we choose the category corresponding to the largest element in $w$ if it is greater than 0, and choose the reference category if the largest element in $w$ is smaller than 0.

\section{Simulations and Real Data Analysis on the QASC}
We evaluated the performance of the proposed model through two simulation studies: (i) simulated artificial data with missing values and (ii) the QASC data set with randomly deleted records. We compared the proposed imputation model with other commonly used procedures to treat missing data.

\subsection{Simulation Based on Artificial Data}
We generated 100 complete multilevel data sets with correlated variables of different types, and then deleted some entries under the MAR assumption. The total number of clusters in each data set was 20, the cluster size was 50, and the five variables $X_1, X_2, X_3, X_4, X_5$ had Gamma, binary, nominal, ordinal, and normal distributions respectively. The variable $X_1$ followed a skewed Gamma distribution: $X_1 \sim Gamma(3,0.5)$. We assumed all the subsequent variables were generated depending on the previous ones, to introduce correlation among variables. Specifically, $X_2$ was a binary variable such that $logit(p_{X_2})=X_1+\epsilon_{ij}$, where $p_{X_2}$ is the probability that $X_2$ equals 1 and $\epsilon_{ij} \sim N(0,1)$. The nominal variable $X_3$ had 4 categories and was generated by a multinomial probit model, so that 3 latent variables were needed: $(l_{X_3,1},l_{X_3,2},l_{X_3,3}) \sim N((X_1,x_2)B_{X_3}, C_{X_3})$, where $B_{X_3}$ is a randomly generated coefficient matrix of dimension $2\times3$ and $C_{X_3}$ is a correlation matrix of dimension $3 \times 3$. The category in $X_3$ was chosen to be $k$ (for k=1,2,3) if $l_{X_3,k}$ was the largest component and was greater than 0; and was chosen to be 4 if $max(l_{X_3})<0$. Because we aimed to create a data set with a multilevel structure, we let the ordinal variable $X_4$ be generated from a random intercept model, $l_{X_4}=b_{X_4,i}+X_1+X_2+ \beta_{X_3} X_3+\epsilon_{ij}$, with $\epsilon_{ij} \sim N(0,1)$, $b_{X_4,i} \sim N(0,\rho_{X_4})$, and $\beta_{X_3}$ a vector of length 3, corresponding to the 3 categories in $X_3$. Three thresholds were used to determine four levels, they were the 20\%, 30\%, 50\% quantiles of $l_{X_4}$. Lastly, the normally distributed variable $X_5$ was also generated from a random intercept model, $X_5=b_{X_5,i}+X_1+X_2+ \beta_{X_3} X_3+\beta_{X_4} x_4+\epsilon_{ij}$, where $\epsilon_{ij} \sim N(0,1)$, $b_{X_5,i} \sim N(0,\rho_{X_5})$, and $\beta_{X_3}$ and $\beta_{X_4}$ are vectors of length 3.

To create missing data under the MAR assumption, we assumed $X_5$ was completely observed and that the probabilities of missingness in $X_j$ ($j=1,...,4$) depended on $X_5$. Specifically, let $p_{mis,ij}$ be the probability that observation $i$ is missing its value for the $X_j$ variable and we assumed that $logit(p_{mis,ij})=\alpha_jX_5$. By adjusting the parameters $\alpha_j$, we can control the missingness in each variable.

We varied the parameters that generated the data to consider different scenarios: (1) missing rates for each variable from low (10\%), median (30\%) to high (50\%); (2) clustering effect from low ($\rho_{X_4}=\rho_{X_5}=0.2$) to high ($\rho_{X_4}=\rho_{X_5}=1$), corresponding to intra-class correlation coefficients of 0.17 and 0.5 respectively. In the imputation step, we set the number of imputations to be $M=10$ \citep{graham2007many}.

\subsection{Simulation Results Summary}
To compare the performance of the proposed method with others, we considered six competing methods, some of which have already been implemented in some software packages. These methods are listed in Table 2. We used the package \textit{mitools} in R \citep{lumley2014mitools} to implement the combining rules (\ref{combining rule}) after $M$ complete data sets had been generated.

\begin{table}
{\renewcommand{\arraystretch}{2}
\scalebox{0.8}{
\begin{tabular}{|c|c|c|c|}
\hline
Method &Description & Software Package  \\
\hline\hline
\parbox[t][3em][t]{5cm}{Complete Case Analysis (\textit{Cluster CC})}&\parbox[t][3em][t]{6cm}{Fits an analyst's model by using the fully observed cases only.}& NA\\
\hline
\parbox[t][4em][t]{5cm}{Joint modelling ignoring clustering effects \\(\textit{JM})}&\parbox[t][4em][t]{6cm}{A multivariate Gaussian distribution is used to approximate the joint distribution of data.}& \textit{Amelia} \citep{honaker2011amelia}\\
\hline
\parbox[t][7em][t]{5cm}{Fully conditional specification ignoring clustering effects (\textit{FCS})}&\parbox[t][7em][t]{6cm}{The sequential method fits generalized linear models to each of the variables with missing values and iterates among these variables to approximate the joint distribution.}&\textit{mi} \citep{su2011multiple}\\
\hline
\parbox[t][7em][t]{5cm}{Joint modelling with clustering effects \\(\textit{Cluster JM})}&\parbox[t][7em][t]{6cm}{A multivariate Gaussian distribution is specified for all the variables with missing values, regressed against the completely observed variables as covariates with random effects.}&\textit{pan} \citep{schafer2002computational}\\
\hline
\parbox[t][5em][t]{5cm}{Fully conditional specification with clustering effects (\textit{Cluster FCS})}&\parbox[t][4em][t]{6cm}{Adds random effects to each of the univariate regression models in the fully conditional specification method.}&\textit{lme4} \citep{bates2014fitting}\\
\hline
\parbox[t][3em][t]{5cm}{Copula model ignoring clustering effects \\(\textit{Copula\_Hoff})}&\parbox[t][4.5em][t]{6cm}{Fits the extended rank likelihood copula.}&\textit{sbgcop} \citep{hoff2007extending}\\
\hline
\parbox[t][3em][t]{5cm}{Copula model with clustering effects \\(\textit{Cluster Copula})}&\parbox[t][4.5em][t]{6cm}{Our proposed method.} &See supplementary materials. \\
\hline
\end{tabular}}}
\caption {Summary of different methods to handle missing data used in simulations.}
\end{table}

The assessment of the relative performance of each method was based on the comparison of the imputation accuracy as well as the 95\% coverage rates of the coefficients in the following random intercept logistic regression as a model of interest. We chose this model arbitrarily, and believe that the results would also hold for other models of interest.

For each of the 100 simulated complete data sets, we fitted the model $logit(p(X_2=1))=b_i+ \beta_0+ \beta_1 X_1+ \beta_2 X_{3,2}+ \beta_3 X_{3,3}+ \beta_4 X_{3,4}, b_i \sim N(0,\sigma^2)$. We used the glmer() function in the \textit{lme4} package in R to obtain parameter estimates for $\beta=(\beta_0,\beta_1,\beta_2,\beta_3,\beta_4)$. These are our `true' parameter values. After deletion of records by MAR, we applied each of the missing data methods listed in Table 2, and calculated point and variance estimates for $\beta$, using the combining rules. We reported the average of the squared bias of the coefficient estimates over the 100 data sets as well as the coverage rates of 95\% confidence intervals.

\rotatebox{90}{\parbox{1\textheight}{
\scalebox{0.72}{
\begin{tabular}{|l|llllllllllllllllllllll}
\hline
\multicolumn{2}{|c|}{\multirow{2}{*}{ICC=0.17}} & \multicolumn{3}{l|}{CC}                                                           & \multicolumn{3}{l|}{JM}                                                       & \multicolumn{3}{l|}{FCS}                                                           & \multicolumn{3}{l|}{Cluster JM}                                                          & \multicolumn{3}{l|}{Cluster FCS}                                                      & \multicolumn{3}{l|}{Copula\_Hoff}                                                       & \multicolumn{3}{l|}{Cluster Copula}                                                      \\ \cline{3-23}
\multicolumn{2}{|c|}{}                          & 10\% & 30\% & \multicolumn{1}{l|}{50\%} & 10\% & 30\% & \multicolumn{1}{l|}{50\%} &10\% & 30\% & \multicolumn{1}{l|}{50\%} & 10\% & 30\% & \multicolumn{1}{l|}{50\%} & 10\% & 30\% & \multicolumn{1}{l|}{50\%} & 10\% & 30\% & \multicolumn{1}{l|}{50\%} & 10\% & 30\% & \multicolumn{1}{l|}{50\%} \\ \hline
\multirow{5}{*}{Bias}        & \multicolumn{1}{l|}{$\beta_0$}       & 0.033                     & 0.146                     & \multicolumn{1}{l|}{0.552}                     & 0.037                     & 0.129                     & \multicolumn{1}{l|}{0.388}                     & 0.014                     & 0.067                     & \multicolumn{1}{l|}{0.198}                     & 0.041                     & 0.125                     & \multicolumn{1}{l|}{0.339}                     & 0.011                     & 0.040                     & \multicolumn{1}{l|}{0.096}                     & 0.014                     & 0.055                     & \multicolumn{1}{l|}{0.104}                    & 0.013                     & 0.054                     & \multicolumn{1}{l|}{0.092}                     \\
                             & \multicolumn{1}{l|}{$\beta_1$}        & 0.040                     & 0.183                     & \multicolumn{1}{l|}{0.309}                     & 0.044                     & 0.133                     & \multicolumn{1}{l|}{0.308}                     & 0.024                     & 0.092                     & \multicolumn{1}{l|}{0.186}                     & 0.022                     & 0.088                     & \multicolumn{1}{l|}{0.399}                     & 0.026                     & 0.105                     & \multicolumn{1}{l|}{0.244}                     & 0.029                     & 0.104                     & \multicolumn{1}{l|}{0.316}                     & 0.026                     & 0.095                     & \multicolumn{1}{l|}{0.187}                     \\
                             & \multicolumn{1}{l|}{$\beta_2$}        & 0.034                     & 0.076                     & \multicolumn{1}{l|}{0.190}                     & 0.026                     & 0.050                     & \multicolumn{1}{l|}{0.104}                     & 0.022                     & 0.055                     & \multicolumn{1}{l|}{0.106}                     & 0.033                     & 0.055                     & \multicolumn{1}{l|}{0.183}                     & 0.021                     & 0.043                     & \multicolumn{1}{l|}{0.099}                     & 0.024                     & 0.053                     & \multicolumn{1}{l|}{0.080}                     & 0.021                     & 0.049                     & \multicolumn{1}{l|}{0.086}                     \\
                             & \multicolumn{1}{l|}{$\beta_3$}        & 0.018                     & 0.073                     & \multicolumn{1}{l|}{0.164}                     & 0.016                     & 0.048                     & \multicolumn{1}{l|}{0.155}                     & 0.010                     & 0.056                     & \multicolumn{1}{l|}{0.118}                     & 0.021                     & 0.051                     & \multicolumn{1}{l|}{0.140}                     & 0.011                     & 0.038                     & \multicolumn{1}{l|}{0.079}                     & 0.012                     & 0.046                     & \multicolumn{1}{l|}{0.080}                     & 0.013                     & 0.051                     & \multicolumn{1}{l|}{0.087}                     \\
                             & \multicolumn{1}{l|}{$\beta_4$}        & 0.045                     & 0.126                     & \multicolumn{1}{l|}{0.373}                     & 0.032                     & 0.194                     & \multicolumn{1}{l|}{0.442}                     & 0.017                     & 0.098                     & \multicolumn{1}{l|}{0.261}                     & 0.020                     & 0.089                     & \multicolumn{1}{l|}{0.383}                     & 0.011                     & 0.047                     & \multicolumn{1}{l|}{0.106}                     & 0.013                     & 0.057                     & \multicolumn{1}{l|}{0.131}                     & 0.012                     & 0.055                     & \multicolumn{1}{l|}{0.122}                     \\ \cline{1-1}
\hline
\hline
\cline{1-1}
\multirow{5}{*}{Coverage} & \multicolumn{1}{l|}{$\beta_0$} & 90  & 77 & \multicolumn{1}{l|}{67} & 87 & 78 & \multicolumn{1}{l|}{79} & 90  & 83  & \multicolumn{1}{l|}{83} & 89  & 76 & \multicolumn{1}{l|}{73} & 93  & 89 & \multicolumn{1}{l|}{87} & 90  & 86  & \multicolumn{1}{l|}{80} & 92  & 90  & \multicolumn{1}{l|}{87} \\
                          & \multicolumn{1}{l|}{$\beta_1$} & 97  & 84 & \multicolumn{1}{l|}{77} & 95 & 90 & \multicolumn{1}{l|}{79} & 98 & 92  & \multicolumn{1}{l|}{81} & 98 & 94 & \multicolumn{1}{l|}{77} & 94 & 90 & \multicolumn{1}{l|}{81} & 89  & 85  & \multicolumn{1}{l|}{79} & 100 & 96  & \multicolumn{1}{l|}{84} \\
                          & \multicolumn{1}{l|}{$\beta_2$} & 100 & 92 & \multicolumn{1}{l|}{87} & 98 & 92 & \multicolumn{1}{l|}{91} & 95  & 100 & \multicolumn{1}{l|}{96} & 97  & 89 & \multicolumn{1}{l|}{88} & 100 & 98 & \multicolumn{1}{l|}{98} & 100 & 100 & \multicolumn{1}{l|}{99} & 99  & 100 & \multicolumn{1}{l|}{98} \\
                          & \multicolumn{1}{l|}{$\beta_3$} & 95  & 94 & \multicolumn{1}{l|}{82} & 93 & 94 & \multicolumn{1}{l|}{88} & 93  & 95  & \multicolumn{1}{l|}{88} & 91  & 90 & \multicolumn{1}{l|}{76} & 91  & 91 & \multicolumn{1}{l|}{85} & 94  & 93  & \multicolumn{1}{l|}{87} & 91  & 92  & \multicolumn{1}{l|}{92} \\
                          & \multicolumn{1}{l|}{$\beta_4$} & 87  & 80 & \multicolumn{1}{l|}{63} & 89 & 76 & \multicolumn{1}{l|}{69} & 93  & 90  & \multicolumn{1}{l|}{83} & 90  & 82 & \multicolumn{1}{l|}{79} & 90  & 89 & \multicolumn{1}{l|}{87} & 95  & 87  & \multicolumn{1}{l|}{89} & 93  & 88  & \multicolumn{1}{l|}{85} \\ \cline{1-1}
\hline
\end{tabular}}
\captionof{table}{A comparison of squared bias and coverage of coefficient estimates of the model of interest under seven methods to handle missing data, with ICC=0.17 and missing rates=10\%, 30\% and 50\%.}
\label{table_exp_settings}
}}
\hspace{2cm}
\rotatebox{90}{\parbox{1\textheight}{
\scalebox{0.72}{
\begin{tabular}{|l|llllllllllllllllllllll}
\hline
\multicolumn{2}{|c|}{\multirow{2}{*}{ICC=0.5}} & \multicolumn{3}{l|}{CC}                                                           & \multicolumn{3}{l|}{JM}                                                       & \multicolumn{3}{l|}{FCS}                                                           & \multicolumn{3}{l|}{Cluster JM}                                                          & \multicolumn{3}{l|}{Cluster FCS}                                                      & \multicolumn{3}{l|}{Copula Hoff}                                                       & \multicolumn{3}{l|}{Cluster Copula}                                                      \\ \cline{3-23}
\multicolumn{2}{|c|}{}                         & 10\%   & 30\%   & \multicolumn{1}{l|}{50\%}  & 10\%    & 30\%    & \multicolumn{1}{l|}{50\%}    & 10\%   & 30\%   & \multicolumn{1}{l|}{50\%}  & 10\%   & 30\%   & \multicolumn{1}{l|}{50\%}   & 10\%     & 30\%    & \multicolumn{1}{l|}{50\%}    & 10\%    & 30\%    & \multicolumn{1}{l|}{50\%}    & 10\%     & 30\%    & \multicolumn{1}{l|}{50\%}    \\ \hline
\multirow{5}{*}{Bias}        & \multicolumn{1}{l|}{$\beta_0$}        & 0.051  & 0.164  & \multicolumn{1}{l|}{0.539} & 0.056   & 0.160   & \multicolumn{1}{l|}{0.481}   & 0.029  & 0.114  & \multicolumn{1}{l|}{0.216} & 0.053  & 0.197  & \multicolumn{1}{l|}{0.411}  & 0.025    & 0.074   & \multicolumn{1}{l|}{0.129}   & 0.030   & 0.100   & \multicolumn{1}{l|}{0.139}   & 0.024    & 0.076   & \multicolumn{1}{l|}{0.127}   \\
                             & \multicolumn{1}{l|}{$\beta_1$}        & 0.035  & 0.210  & \multicolumn{1}{l|}{0.542} & 0.049   & 0.143   & \multicolumn{1}{l|}{0.424}   & 0.019  & 0.084  & \multicolumn{1}{l|}{0.212} & 0.029  & 0.102  & \multicolumn{1}{l|}{0.373}  & 0.022    & 0.136   & \multicolumn{1}{l|}{0.303}   & 0.036   & 0.191   & \multicolumn{1}{l|}{0.364}   & 0.031    & 0.183   & \multicolumn{1}{l|}{0.290}   \\
                             & \multicolumn{1}{l|}{$\beta_2$}        & 0.040  & 0.098  & \multicolumn{1}{l|}{0.235} & 0.068   & 0.180   & \multicolumn{1}{l|}{0.327}   & 0.027  & 0.070  & \multicolumn{1}{l|}{0.122} & 0.034  & 0.173  & \multicolumn{1}{l|}{0.391}  & 0.023    & 0.069   & \multicolumn{1}{l|}{0.111}   & 0.028   & 0.077   & \multicolumn{1}{l|}{0.093}   & 0.022    & 0.060   & \multicolumn{1}{l|}{0.094}   \\
                             & \multicolumn{1}{l|}{$\beta_3$}        & 0.035  & 0.105  & \multicolumn{1}{l|}{0.260} & 0.115   & 0.250   & \multicolumn{1}{l|}{0.250}   & 0.024  & 0.088  & \multicolumn{1}{l|}{0.162} & 0.035  & 0.119  & \multicolumn{1}{l|}{0.221}  & 0.024    & 0.071   & \multicolumn{1}{l|}{0.108}   & 0.023   & 0.073   & \multicolumn{1}{l|}{0.127}   & 0.024    & 0.071   & \multicolumn{1}{l|}{0.103}   \\
                             & \multicolumn{1}{l|}{$\beta_4$}        & 0.089  & 0.303  & \multicolumn{1}{l|}{0.492} & 0.098   & 0.319   & \multicolumn{1}{l|}{0.514}   & 0.044  & 0.200  & \multicolumn{1}{l|}{0.340} & 0.045  & 0.196  & \multicolumn{1}{l|}{0.456}  & 0.038    & 0.138   & \multicolumn{1}{l|}{0.175}   & 0.048   & 0.178   & \multicolumn{1}{l|}{0.211}   & 0.040    & 0.138   & \multicolumn{1}{l|}{0.183}   \\
\hline
\hline
\cline{1-1}
\multirow{5}{*}{Coverage} & \multicolumn{1}{l|}{$\beta_0$} & 91  & 76 & \multicolumn{1}{l|}{74} & 87 & 80 & \multicolumn{1}{l|}{78} & 89  & 80 & \multicolumn{1}{l|}{79} & 88  & 89 & \multicolumn{1}{l|}{90} & 90  & 85 & \multicolumn{1}{l|}{88} & 90  & 89 & \multicolumn{1}{l|}{85} & 92  & 91 & \multicolumn{1}{l|}{91} \\
                          & \multicolumn{1}{l|}{$\beta_1$} & 92  & 80 & \multicolumn{1}{l|}{65} & 94 & 86 & \multicolumn{1}{l|}{79} & 93  & 92 & \multicolumn{1}{l|}{78} & 96  & 92 & \multicolumn{1}{l|}{73} & 95  & 90 & \multicolumn{1}{l|}{74} & 88  & 83 & \multicolumn{1}{l|}{78} & 95  & 93 & \multicolumn{1}{l|}{80} \\
                          & \multicolumn{1}{l|}{$\beta_2$}& 100 & 92 & \multicolumn{1}{l|}{80} & 97 & 91 & \multicolumn{1}{l|}{80} & 95  & 96 & \multicolumn{1}{l|}{92} & 94  & 90 & \multicolumn{1}{l|}{86} & 98 & 98 & \multicolumn{1}{l|}{93} & 100 & 97 & \multicolumn{1}{l|}{97} & 100 & 99 & \multicolumn{1}{l|}{91} \\
                          & \multicolumn{1}{l|}{$\beta_3$} & 89  & 89 & \multicolumn{1}{l|}{82} & 90 & 76 & \multicolumn{1}{l|}{74} & 90 & 95 & \multicolumn{1}{l|}{86} & 90 & 88 & \multicolumn{1}{l|}{78} & 88  & 87 & \multicolumn{1}{l|}{87} & 90  & 87 & \multicolumn{1}{l|}{87} & 89  & 92 & \multicolumn{1}{l|}{90} \\
                          & \multicolumn{1}{l|}{$\beta_4$} & 88  & 76 & \multicolumn{1}{l|}{66} & 87 & 81 & \multicolumn{1}{l|}{70} & 92  & 84 & \multicolumn{1}{l|}{73} & 90 & 83 & \multicolumn{1}{l|}{78} & 91  & 88 & \multicolumn{1}{l|}{82} & 90  & 82 & \multicolumn{1}{l|}{84} & 90  & 84 & \multicolumn{1}{l|}{82} \\ \cline{1-1}
\hline
\end{tabular}}
\captionof{table}{A comparison of squared bias and coverage of coefficient estimates of the model of interest under seven methods to handle missing data, with ICC=0.5 and missing rates=10\%, 30\% and 50\%.}
\label{table_exp_settings}
}}

Table 3 summarizes the results of the simulation experiments under the three missingness rates (10\%, 30\% and 50\%) using the seven methods, when the ICC used to generate the variables $X_4$ and $X_5$ is 0.17. When the missingness rate is 10\%, all the approaches give reasonably good results in terms of achieving the nominal coverage rate - 95\%, though \textit{CC} and the two joint modelling approaches (\textit{JM} and \textit{Cluster JM}) do worse than the others. A possible reason for this is that in the joint modelling approaches, multivariate Gaussian distributions were specified and this is clearly not true in our data generating process, whereas in the sequential imputation approaches (\textit{FCS} and \textit{Cluster FCS}) more flexible univariate imputation models were allowed to best accommodate different variable types. For the copula-based methods, the empirical distribution function transformations were applied before fitting a multivariate Gaussian distribution on the latent variable scale where the dependence among the variables was captured. In addition, the squared bias increases with an increase in missingness rate as expected. With a moderate to high level of missingness, \textit{Cluster FCS} and our proposed method (\textit{Cluster Copula}) tend to outperform \textit{FCS} and \textit{Copula\_Hoff}. While all the methods suffer from under-coverage when the missing rates are 30\% and 50\%, \textit{CC} seems to be the worst, producing the most biased results. The results meet our expectation because as the percentage of missing data increases, there is less observed data available to capture the complex dependency among variables. Under the MAR assumption, \textit{CC} causes the most biased results by only using the complete records while its alternatives make use of all the observed data.

Table 4 is similar to Table 3 except that the performance is evaluated at ICC=0.5. In other words, the data sets exhibit higher levels of clustering. Compared to the results in Table 3, the results are worse across all methods for the higher ICC value. The imputation methods which take into account clustering effects almost always do better than their counterparts, which is not that obvious in Table 3 when ICC=0.17. Conditional imputation methods do better than joint modelling approaches, and the two copula-based methods tend to achieve the best results, for almost half of the simulation settings with the smallest squared bias.

We also compared the imputation accuracy. That is, for each data value we calculated the discrepancy between the average of the 10 imputed values and the before-deletion true values. Note that this comparison is not applied to the \textit{CC} method.  The Euclidean distance was used to measure the imputation accuracy in the continuous variable $X_1$ and the ordinal variable $X_3$: $\frac{1}{10}\sum_{m=1}^{10} \frac{\sum_{i=1}^N (X_{i,true}-X_{i,imp}^{(m)})^2}{\#miss~X}$, and the misclassification rate was used to measure the imputation accuracy in the binary variable $X_2$ and the nominal variable $X_4$: $1-\frac{1}{10}\sum_{m=1}^{10}\frac{\sum_{i=1}^N \mathbbm{1}_{(X_{i,true}=X_{i,imp}^{(m)})}}{\#miss~X}$.

Figure 1 shows the results of the imputation accuracy for each simulation study. The points are the means of the Euclidian distances/misclassification rates over the missing observations in a single data set, and the error bars show the $5\%$ and $95\%$ quantiles over the 100 data sets. For variable $X_1$ which follows a Gamma distribution, there is not much difference in imputation accuracy over the six methods. For the nominal variable $X_3$ our proposed \textit{Cluster Copula} method always performs the best except for the top-left panel, while the \textit{JM} approach is always the worst. The misclassification rates for the binary variable $X_2$ are smallest in all the scenarios when using our proposed \textit{Cluster Copula} model but do not differ much from those of the other methods. The misclassification rates for the ordinal variable $X_4$ are again highest for \textit{JM} and the rates for the copula-based methods are smaller than the others when the missing rates are 30\% and 50\%. Generally speaking, the copula based methods tend to impute more accurately for categorical variables but also do no worse than other methods for continuous variables. The joint modelling methods, especially \textit{JM}, give the least accurate imputation as the multivariate Gaussian distribution assumption does not hold. As the missingness rate and/or ICC increase, all the methods for every variable perform comparatively worse in terms of having a larger disparity compared with the true values and higher misclassification rates, but the patterns of relative performance between the six methods remains broadly the same.

\begin{figure}
\includegraphics[width=16.5cm, height=18cm]{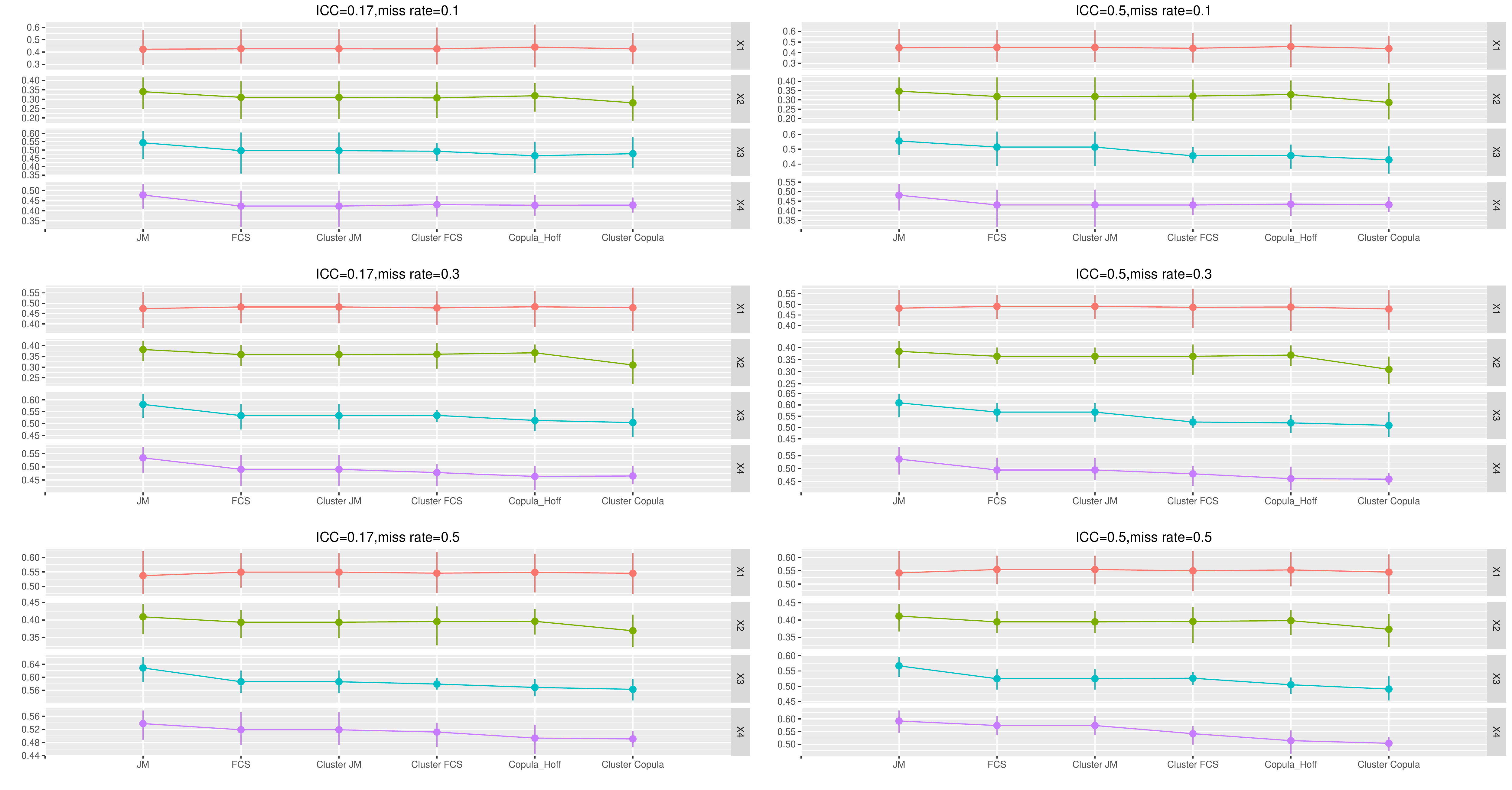}
\caption{Prediction accuracy of the simulated data sets, with the points stand for the means of the Euclidian distances/misclassification rates, and the error bars stand for the $5\% - 95\%$ quantiles over the 100 data sets.}
\end{figure}

\subsection{Simulation Based on the QASC Data Set}
We also ran simulation studies using the QASC data set to evaluate our proposed method and other competing methods. Here we treated all the complete cases in the QASC data set (75.34\% of the original data set) as the `true' data, and sub-sampled 300 patients, 100 times to create 100 sub data sets. Then for each of the sub data sets, missing values were created, trying to mimic the missing data pattern in the original data set. We distinguish between the demographic variables which we treat as MCAR and the process of care variables which we treat as MAR. Specifically, for the demographic variables: `ATSI', `age', `education' and `marital status', values were randomly deleted to roughly match the missingness percentages in Table 1. For the process of care variables and outcome variables, we assumed their missingness depended on the completely observed variables. A missing indicator variable was associated with every variable with missing data which equaled 1 if an entry was missing. For the missing indicators, we fitted logistic regression models on the original data set for `time taken to hospital', `mean temperature', `modified Rankin Scale', `Bartell Index', `physical health score' and `mental health score' respectively against `gender', `period' and `treatment', and the probabilities of missingness for the sub-sampled data sets were decided by the predicted values of these logistic regression models. We noticed that 9.39\% of `Bartell Index', `physical health score' and `mental health score' were missing together, and we also took this into account when creating missing data.

The relative performance for each method was also compared based on the average imputation accuracy and the squared bias and 95\% coverage rate of interval estimates of parameters for some models of interest. Ten imputations were created for all the six imputation methods. The accuracy is shown in Table 5. All the discrepancies between the imputed values and the true values were measured by Euclidean distance except for the nominal variables `marital status' and `ATSI' which used the misclassification rates. Our proposed imputation model achieves the smallest disparity more than half of the time (7/11) and \textit{Copula\_Hoff} is superior in performance to the other four methods. It is interesting to note that joint modelling methods perform better than their FCS counterparts (\textit{JM} vs.\textit{FCS} and \textit{Cluster JM} vs.\textit{Cluster FCS}) and adding clustering effects enhances the imputation accuracy.

\begin{table}
\centering
\caption{Imputation accuracy in QASC with randomly deleted records, measured by the average Euclidean distances between the imputed values and the true values for the first nine variables and misclassification rate for the last variable.}
\scalebox{0.9}{
\begin{tabular}{@{}lllllll@{}}
\toprule
    Variable       & JM  & FCS      & Cluster JM            & Cluster FCS          & Copula\_Hoff           & Cluster Copula          \\ \midrule
time taken to hospital & 349.87 & 368.66 & 144.57 & 224.46  & 145.82 & \textbf{135.22} \\
education  & \textbf{2.69}   & 4.3    & 2.74   & 3.79   & 2.79   & \textbf{2.69}            \\
age        & 336.37 & 511.35 & 301.54 & \textbf{249.43} & 290.48 & 254.84 \\
modified Rankin Scale        & 2.78   & 5.12   & 2.84   & 4.16   & 2.87   & \textbf{2.72}            \\
Bartell Index         & 535.38 & 849.1  & 594.52 & 654.67 & 548.47 & \textbf{443.1} \\
physical health score       & 163.23 & 302.81 & 177.68 & 186.11 & 171.54 & \textbf{161.42}          \\
mental health score      & 286.34 & 451.16 & 260.27 & 344.57 & 253.36 & \textbf{241.71} \\
length of stay        & \textbf{126.4}  & 350    & 172.88 & 196.01 & 169.54 & 158.15          \\
mean temperature   & 0.19   & 0.27   & 1.83  & 0.19   & 0.14   & \textbf{0.13}   \\
marital status   & 0.51   & 0.61   & 0.53   & 0.55   & \textbf{0.46}   & 0.5           \\
ATSI &\textbf{0.014}  & 0.122  & 0.015  & 0.02   & 0.027  & 0.024 \\ \bottomrule
\end{tabular}}
\end{table}

The models of interest are based on the models fitted in \cite{middleton2011implementation}. They fitted logistic regression models for the dichotomous outcomes - `Bartell Index' with cut points equal to 60 and 95, and `modified Rankin Scale' with cut point equaled to 2; and linear models for the continuous variables `physical health score' and `mental health score', including as predictors the variables `treatment', `period' and the interaction between `treatment' and `period'. The models are

\begin{equation*}\label{Interaction QASC}
  \begin{aligned}
  &\mbox{logit}(mrs2)=b_i+\beta_0+\beta_1 period+ \beta_2 treatment +\beta_3 treatment*period,\\
  &\mbox{logit}(bi60)=b_i+\beta_0+\beta_1 period+ \beta_2 treatment +\beta_3 treatment*period,\\
  &\mbox{logit}(bi90)=b_i+\beta_0+\beta_1 period+ \beta_2 treatment +\beta_3 treatment*period,\\
  &mcs=b_i+\beta_0+\beta_1 period+ \beta_2 treatment +\beta_3 treatment*period+\epsilon,\\
  &pcs=b_i+\beta_0+\beta_1 period+ \beta_2 treatment +\beta_3 treatment*period+\epsilon.
  \end{aligned}
\end{equation*}
The coefficient $\beta_3$ and its p-value were used to see if the pre-post change in the intervention group was statistically significant to the change in the control group.  All the models included a random intercept term, $b_i$, to capture the clustering effects.

We first fitted the five models of interest on the completely observed patients in each of the 100 sub data sets, and obtained the parameter estimates $\beta=(\beta_0,\beta_1,\beta_2,\beta_3)$ and treated them as the true values. Then the parameter estimates from all the seven competing methods were compared against the true parameters, and the 95\% coverage rates were obtained from the 100 repetitions. The results are reported in Table 6. The \textit{CC} approach has the largest bias and the smallest coverage rate. This is not unexpected because the missing data were generated under the MAR assumption and by \textit{CC} only about 40\% of the data were used to fit the models so that the coefficient estimates are biased with large uncertainty. The proposed method \textit{Cluster Copula} and \textit{Copula\_Hoff} outperform the other methods with \textit{Copula\_Hoff} doing marginally better than \textit{Cluster Copula} for the first and second logistic models `mrs2' and `bi60', and \textit{Cluster Copula} doing better for the fifth linear model for `pcs'. There is little difference between the two copula based methods, because the clustering effects were small in the QASC data set (ICC in the models of interest lay between 0.009 and 0.026), and only one nominal variable (marital status) was considered in the imputation models but did not enter into the models of interest later. In other words, taking the clustering effect into account and giving special treatment to the nominal variable does not affect the inference too much in this case. However, we do observe that when ICC is higher in the variable `pcs', our proposed model achieves better imputation accuracy.

\rotatebox{90}{\parbox{1\textheight}{
		\scalebox{0.6}{
			\begin{tabular}{|l|l|lllllllllllllllllllll}
				\hline
				\multicolumn{2}{|l|}{\multirow{2}{*}{}} & \multicolumn{3}{l|}{CC}& \multicolumn{3}{l|}{JM}& \multicolumn{3}{l|}{FCS}& \multicolumn{3}{l|}{Cluster JM} & \multicolumn{3}{l|}{Cluster FCS}                                                        & \multicolumn{3}{l|}{Copula\_Hoff}& \multicolumn{3}{l|}{Cluster Copula}  \\ \cline{3-23}
				\multicolumn{2}{|l|}{}                  & Bias & SD & \multicolumn{1}{l|}{Coverage} & Bias & sd & \multicolumn{1}{l|}{Coverage} & Bias & SD & \multicolumn{1}{l|}{Coverage} & Bias & SD & \multicolumn{1}{l|}{Coverage} & Bias & sd & \multicolumn{1}{l|}{Coverage} & Bias & SD & \multicolumn{1}{l|}{Coverage} & Bias & SD & \multicolumn{1}{l|}{Coverage} \\ \hline
				\multirow{4}{*}{Modified Rankin Scale 2}    & \multicolumn{1}{l|}{$\beta_0$}    &0.029 & 0.426 & \multicolumn{1}{l|}{91} & 0.007 & 0.282 & \multicolumn{1}{l|}{91}  & 0.013 & 0.284 & \multicolumn{1}{l|}{92} & 0.006 & 0.283 & \multicolumn{1}{l|}{90} & 0.01  & 0.274 & \multicolumn{1}{l|}{92}  & 0.006 & 0.282 & \multicolumn{1}{l|}{92}  & 0.007 & 0.285 & \multicolumn{1}{l|}{92} \\
				\multicolumn{1}{|l|}{} & \multicolumn{1}{l|}{$\beta_1$}&0.049 & 0.671 & \multicolumn{1}{l|}{94} & 0.017 & 0.43  & \multicolumn{1}{l|}{96}  & 0.021 & 0.431 & \multicolumn{1}{l|}{96} & 0.015 & 0.433 & \multicolumn{1}{l|}{98} & 0.019 & 0.409 & \multicolumn{1}{l|}{94}  & 0.013 & 0.432 & \multicolumn{1}{l|}{98}  & 0.016 & 0.434 & \multicolumn{1}{l|}{97} \\
				\multicolumn{1}{|l|}{} & \multicolumn{1}{l|}{$\beta_2$}&0.022 & 0.534 & \multicolumn{1}{l|}{93} & 0.013 & 0.344 & \multicolumn{1}{l|}{100} & 0.014 & 0.344 & \multicolumn{1}{l|}{99} & 0.014 & 0.343 & \multicolumn{1}{l|}{98} & 0.014 & 0.391 & \multicolumn{1}{l|}{100} & 0.011 & 0.345 & \multicolumn{1}{l|}{100} & 0.013 & 0.344 & \multicolumn{1}{l|}{99} \\
				\multicolumn{1}{|l|}{} & \multicolumn{1}{l|}{$\beta_3$}&0.043 & 0.827 & \multicolumn{1}{l|}{88} & 0.026 & 0.521 & \multicolumn{1}{l|}{91}  & 0.033 & 0.523 & \multicolumn{1}{l|}{89} & 0.023 & 0.52  & \multicolumn{1}{l|}{90} & 0.032 & 0.524 & \multicolumn{1}{l|}{90}  & 0.022 & 0.522 & \multicolumn{1}{l|}{93}  & 0.024 & 0.518 & \multicolumn{1}{l|}{92}\\\cline{1-2}\hline\hline
				
				\multirow{4}{*}{Bartell Index 60}    & \multicolumn{1}{l|}{$\beta_0$}    & 0.051 & 0.584 & \multicolumn{1}{l|}{87} & 0.043 & 0.315 & \multicolumn{1}{l|}{89} & 0.032 & 0.329 & \multicolumn{1}{l|}{90} & 0.048 & 0.316 & \multicolumn{1}{l|}{86} & 0.035 & 0.315 & \multicolumn{1}{l|}{88} & 0.012 & 0.327 & \multicolumn{1}{l|}{91} & 0.021 & 0.333 & \multicolumn{1}{l|}{90} \\
				\multicolumn{1}{|l|}{} & \multicolumn{1}{l|}{$\beta_1$}&0.028 & 0.586 & \multicolumn{1}{l|}{97} & 0.027 & 0.478 & \multicolumn{1}{l|}{95} & 0.024 & 0.487 & \multicolumn{1}{l|}{97} & 0.022 & 0.480 & \multicolumn{1}{l|}{97} & 0.024 & 0.495 & \multicolumn{1}{l|}{95} & 0.020 & 0.495 & \multicolumn{1}{l|}{97} & 0.026 & 0.501 & \multicolumn{1}{l|}{95} \\
				\multicolumn{1}{|l|}{} & \multicolumn{1}{l|}{$\beta_2$}&0.043 & 0.819 & \multicolumn{1}{l|}{85} & 0.035 & 0.395 & \multicolumn{1}{l|}{89} & 0.029 & 0.405 & \multicolumn{1}{l|}{92} & 0.025 & 0.396 & \multicolumn{1}{l|}{93} & 0.030 & 0.398 & \multicolumn{1}{l|}{90} & 0.016 & 0.411 & \multicolumn{1}{l|}{93} & 0.018 & 0.415 & \multicolumn{1}{l|}{93} \\
				\multicolumn{1}{|l|}{} & \multicolumn{1}{l|}{$\beta_3$}&0.083 & 0.746 & \multicolumn{1}{l|}{90} & 0.055 & 0.606 & \multicolumn{1}{l|}{92} & 0.072 & 0.624 & \multicolumn{1}{l|}{87} & 0.046 & 0.610 & \multicolumn{1}{l|}{95} & 0.048 & 0.635 & \multicolumn{1}{l|}{97} & 0.038 & 0.634 & \multicolumn{1}{l|}{98} & 0.055 & 0.643 &\multicolumn{1}{l|}{ 98} \\\cline{1-2}\hline\hline
				
				\multirow{4}{*}{Bartell Index 95}    & \multicolumn{1}{l|}{$\beta_0$}    & 1.228 & 1.293 & \multicolumn{1}{l|}{74} & 0.148 & 0.881 & \multicolumn{1}{l|}{81} & 0.163 & 0.874 & \multicolumn{1}{l|}{80} & 0.145 & 0.875 & \multicolumn{1}{l|}{80} & 0.160 & 0.792 & \multicolumn{1}{l|}{78} & 0.117 & 0.608 & \multicolumn{1}{l|}{85} & 0.123 & 0.627 & \multicolumn{1}{l|}{82} \\
				\multicolumn{1}{|l|}{} & \multicolumn{1}{l|}{$\beta_1$}&3.252 & 1.379 & \multicolumn{1}{l|}{90} & 2.518 & 0.752 & \multicolumn{1}{l|}{96} & 3.061 & 0.654 & \multicolumn{1}{l|}{98} & 1.945 & 0.733 & \multicolumn{1}{l|}{96} & 1.854 & 0.780 & \multicolumn{1}{l|}{95} & 1.770 & 0.627 & \multicolumn{1}{l|}{97} & 1.422 & 0.831 & \multicolumn{1}{l|}{96} \\
				\multicolumn{1}{|l|}{} & \multicolumn{1}{l|}{$\beta_2$}&0.263 & 1.249 & \multicolumn{1}{l|}{72} & 0.168 & 0.971 & \multicolumn{1}{l|}{87} & 0.212 & 0.967 & \multicolumn{1}{l|}{80} & 0.170 & 0.966 & \multicolumn{1}{l|}{78} & 0.162 & 0.814 & \multicolumn{1}{l|}{79} & 0.181 & 0.710 & \multicolumn{1}{l|}{86} & 0.175 & 0.722 & \multicolumn{1}{l|}{87} \\
				\multicolumn{1}{|l|}{} & \multicolumn{1}{l|}{$\beta_3$}&2.953 & 1.092 & \multicolumn{1}{l|}{83} & 2.601 & 0.750 & \multicolumn{1}{l|}{91} & 3.207 & 0.756 & \multicolumn{1}{l|}{85} & 2.043 & 0.733 & \multicolumn{1}{l|}{89} & 1.998 & 0.691 & \multicolumn{1}{l|}{86} & 1.896 & 0.629 & \multicolumn{1}{l|}{93} & 1.599 & 0.780 & \multicolumn{1}{l|}{94}                       \\ \cline{1-2}\hline\hline
				
				\multirow{4}{*}{Mental Health Score}    & \multicolumn{1}{l|}{$\beta_0$}    & 1.791 & 2.398 & \multicolumn{1}{l|}{86} & 1.659 & 2.092 & \multicolumn{1}{l|}{86}  & 0.672 & 1.763 & \multicolumn{1}{l|}{88}  & 0.356 & 1.616 & \multicolumn{1}{l|}{89}  & 0.602 & 1.609 & \multicolumn{1}{l|}{85}  & 0.367 & 1.614 & \multicolumn{1}{l|}{90}  & 0.406 & 1.638 & \multicolumn{1}{l|}{87}  \\
				\multicolumn{1}{|l|}{} & \multicolumn{1}{l|}{$\beta_1$}&2.101 & 3.778 & \multicolumn{1}{l|}{98} & 2.334 & 2.569 & \multicolumn{1}{l|}{100} & 1.178 & 1.587 & \multicolumn{1}{l|}{99}  & 0.892 & 1.498 & \multicolumn{1}{l|}{98}  & 0.952 & 1.487 & \multicolumn{1}{l|}{97}  & 0.913 & 1.495 & \multicolumn{1}{l|}{98}  & 0.968 & 1.515 & \multicolumn{1}{l|}{99}  \\
				\multicolumn{1}{|l|}{} & \multicolumn{1}{l|}{$\beta_2$}&0.951 & 2.984 & \multicolumn{1}{l|}{82} & 1.042 & 2.492 & \multicolumn{1}{l|}{85}  & 0.905 & 2.110 & \multicolumn{1}{l|}{84}  & 0.651 & 2.004 & \multicolumn{1}{l|}{87}  & 0.703 & 1.994 & \multicolumn{1}{l|}{88}  & 0.659 & 2.012 & \multicolumn{1}{l|}{90}  & 0.641 & 1.976 & \multicolumn{1}{l|}{89}  \\
				\multicolumn{1}{|l|}{} & \multicolumn{1}{l|}{$\beta_3$}&1.338 & 3.619 & \multicolumn{1}{l|}{96} & 1.536 & 3.081 & \multicolumn{1}{l|}{100} & 1.459 & 2.164 & \multicolumn{1}{l|}{100} & 1.226 & 2.024 & \multicolumn{1}{l|}{100} & 1.285 & 2.054 & \multicolumn{1}{l|}{100} & 1.199 & 2.058 & \multicolumn{1}{l|}{100} & 1.272 & 1.988 & \multicolumn{1}{l|}{100}   \\ \cline{1-2}\hline\hline
				
				\multirow{4}{*}{Physical Health Score}    & \multicolumn{1}{l|}{$\beta_0$}    & 0.695 & 2.158 & \multicolumn{1}{l|}{85} & 0.708 & 1.819 & \multicolumn{1}{l|}{85} & 0.343 & 1.579 & \multicolumn{1}{l|}{80} & 0.233 & 1.469 & \multicolumn{1}{l|}{86} & 0.309 & 1.452 & \multicolumn{1}{l|}{84} & 0.222 & 1.455 & \multicolumn{1}{l|}{88} & 0.197 & 1.488 & \multicolumn{1}{l|}{90} \\
				\multicolumn{1}{|l|}{} & \multicolumn{1}{l|}{$\beta_1$}&0.871 & 2.385 & \multicolumn{1}{l|}{95} & 0.898 & 1.294 & \multicolumn{1}{l|}{98} & 0.518 & 1.325 & \multicolumn{1}{l|}{97} & 0.457 & 1.247 & \multicolumn{1}{l|}{97} & 0.466 & 1.237 & \multicolumn{1}{l|}{97} & 0.425 & 1.230 & \multicolumn{1}{l|}{98} & 0.390 & 1.258 & \multicolumn{1}{l|}{98} \\
				\multicolumn{1}{|l|}{} & \multicolumn{1}{l|}{$\beta_2$}&0.684 & 2.640 & \multicolumn{1}{l|}{84} & 0.586 & 1.841 & \multicolumn{1}{l|}{88} & 0.378 & 1.823 & \multicolumn{1}{l|}{90} & 0.339 & 1.751 & \multicolumn{1}{l|}{92} & 0.391 & 1.742 & \multicolumn{1}{l|}{91} & 0.350 & 1.755 & \multicolumn{1}{l|}{93} & 0.331 & 1.731 & \multicolumn{1}{l|}{92} \\
				\multicolumn{1}{|l|}{} & \multicolumn{1}{l|}{$\beta_3$}&0.928 & 2.085 & \multicolumn{1}{l|}{87} & 1.139 & 2.079 & \multicolumn{1}{l|}{85} & 0.930 & 1.754 & \multicolumn{1}{l|}{85} & 0.810 & 1.656 & \multicolumn{1}{l|}{90} & 0.964 & 1.801 & \multicolumn{1}{l|}{87} & 0.876 & 1.654 & \multicolumn{1}{l|}{89} & 0.815 & 1.603 & \multicolumn{1}{l|}{92}\\ \cline{1-2}\hline
			\end{tabular}}
			\captionof{table}{A comparison of squared bias in point estimates of coefficients, standard deviations and 95\% coverage of the five models of interest under seven treatments of missing data in the 100 sub sampled QASC data sets.}
			\label{table_exp_settings}
		}}

\subsection{Application to QASC Data Set}
We now apply our proposed method to impute missing data in the original QASC data set with a total of 1480 patients. Unlike in Section 4.3 where we deliberately deleted some records so that we knew the true values, we do not know the true missing values here and therefore cannot measure imputation accuracy. We check the imputation quality by using diagnostics discussed in \cite{abayomi2008diagnostics} and \cite{su2011multiple}.  Specifically, we examined the trace plots of the parameters and convergence in our proposed model (not shown here) and plotted the univariate densities/frequencies of the fully observed values (in black) and the average imputed values (in six colors) for some variables (see Figure 2). All the imputation methods generally agree with the complete data for the continuous variables `length of stay' and `age' and there are small disagreements for the variables `mental health score' and `physical health score'. The imputed values seem to be more spread out for `Bartell Index' than the observed data which is concentrated around 0. Overall, the frequencies of the categorical variables match the observed data with a few exceptions, for example, \textit{FCS} imputes significantly more at level 4 for `Marital status'; and \textit{JM} does not have any imputed values that fall into level 6 for `Modified Rankin Scale'. The departure from the observed data does not necessarily mean the imputation is poor, rather it may mean that the distribution of the missing data is different from what is observed, probably because of the missing data process is MAR rather than MCAR, lack of fit in the imputation model, etc. 

\begin{figure}
\center
\includegraphics[width=17cm, height=15cm]{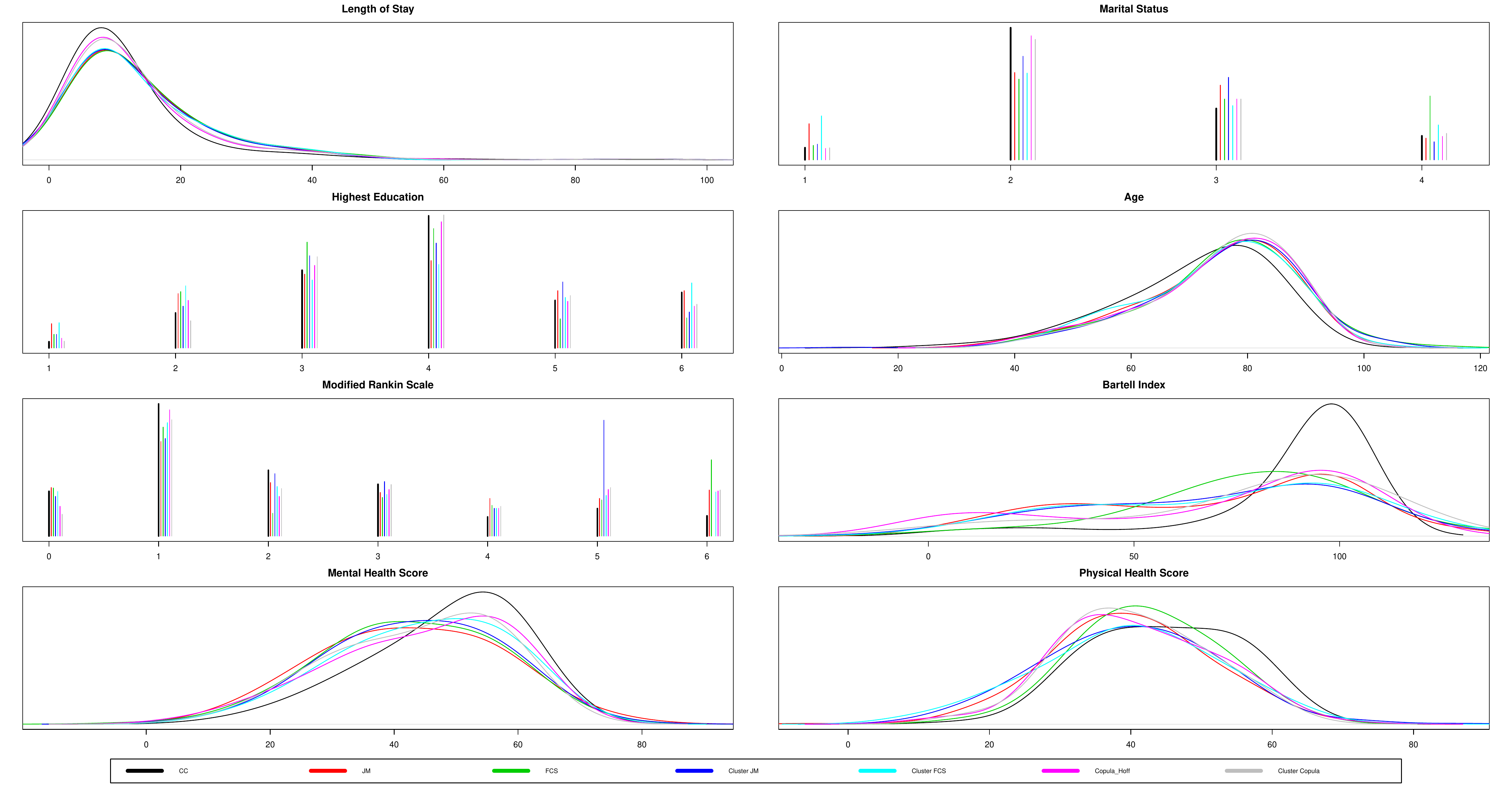}
\caption{Univariate densities/frequencies of the fully observed values (in black) and the average imputed values (in six colors) of eight variables in the original QASC data set.}
\end{figure}

We also report the point estimates of coefficients as well as the standard deviations and p-values of the five models of interest in Table 7, by \textit{CC} and the six imputation methods. While there are differences in the parameter estimates, the p-values across all the methods generally agree with each other, leading to the same clinical conclusions. There are some exceptions, for example in the random intercept logistic regression model for `Bartell Index 60', the coefficient of the interaction term $\beta_3$ is significant at the 0.1 level for the methods \textit{CC}, \textit{FCS} and our proposed \textit{Cluster Copula} method, but significant only at the 0.05 level for the methods \textit{JM}, \textit{Cluster JM}, \textit{Cluster FCS} and \textit{Copula\_Hoff}.

\rotatebox{90}{\parbox{1\textheight}{
\scalebox{0.55}{
\begin{tabular}{|l|l|lllllllllllllllllllll}
\hline
\multicolumn{2}{|l|}{\multirow{2}{*}{}} & \multicolumn{3}{l|}{CC}& \multicolumn{3}{l|}{JM}& \multicolumn{3}{l|}{FCS}& \multicolumn{3}{l|}{Cluster JM} & \multicolumn{3}{l|}{Cluster FCS}                                                        & \multicolumn{3}{l|}{Copula\_Hoff}& \multicolumn{3}{l|}{Cluster Copula}  \\ \cline{3-23}
\multicolumn{2}{|l|}{}                  & point & sd & \multicolumn{1}{l|}{p-value} & point & sd & \multicolumn{1}{l|}{p-value} & point & sd & \multicolumn{1}{l|}{p-value} & point & sd & \multicolumn{1}{l|}{p-value} & point & sd & \multicolumn{1}{l|}{p-value} & point & sd & \multicolumn{1}{l|}{p-value} & point & sd & \multicolumn{1}{l|}{p-value} \\ \hline
\multirow{4}{*}{Modified Rankin Scale 2}    & \multicolumn{1}{l|}{$\beta_0$}    & -0.747                     & 0.156                   & \multicolumn{1}{l|}{\textless.001}                & -0.709                     & 0.152                   & \multicolumn{1}{l|}{\textless.001}                & -0.520                     & 0.147                   & \multicolumn{1}{l|}{\textless.001}                & -0.712                     & 0.160                   & \multicolumn{1}{l|}{\textless.001}                & -0.771                     & 0.150                   & \multicolumn{1}{l|}{\textless.001}                & -0.721                     & 0.151                   & \multicolumn{1}{l|}{\textless.001}                & -0.683                     & 0.154                   & \multicolumn{1}{l|}{\textless.001}                \\
  \multicolumn{1}{|l|}{}                       & \multicolumn{1}{l|}{$\beta_1$}    & 0.114                      & 0.163                   & \multicolumn{1}{l|}{0.483}                        & 0.091                      & 0.161                   & \multicolumn{1}{l|}{0.574}                        & 0.179                      & 0.152                   & \multicolumn{1}{l|}{0.238}                        & 0.102                      & 0.170                   & \multicolumn{1}{l|}{0.549}                        & 0.131                      & 0.163                   & \multicolumn{1}{l|}{0.421}                        & 0.092                      & 0.162                   & \multicolumn{1}{l|}{0.571}                        & 0.125                      & 0.159                   & \multicolumn{1}{l|}{0.434}                        \\
   \multicolumn{1}{|l|}{}                      & \multicolumn{1}{l|}{$\beta_2$}    & 0.241                      & 0.226                   & \multicolumn{1}{l|}{0.285}                        & 0.199                      & 0.218                   & \multicolumn{1}{l|}{0.362}                        & 0.231                      & 0.213                   & \multicolumn{1}{l|}{0.279}                        & 0.199                      & 0.224                   & \multicolumn{1}{l|}{0.375}                        & 0.227                      & 0.216                   & \multicolumn{1}{l|}{0.293}                        & 0.192                      & 0.221                   & \multicolumn{1}{l|}{0.384}                        & 0.143                      & 0.228                   & \multicolumn{1}{l|}{0.530}                        \\
    \multicolumn{1}{|l|}{}                     & \multicolumn{1}{l|}{$\beta_3$}    & -0.659                     & 0.244                   & \multicolumn{1}{l|}{0.007}                        & -0.588                     & 0.237                   & \multicolumn{1}{l|}{0.013}                        & -0.569                     & 0.225                   & \multicolumn{1}{l|}{0.011}                        & -0.601                     & 0.250                   & \multicolumn{1}{l|}{0.016}                        & -0.621                     & 0.242                   & \multicolumn{1}{l|}{0.010}                        & -0.616                     & 0.244                   & 0.012                        & \multicolumn{1}{l|}{-0.604}                     & 0.240                   & \multicolumn{1}{l|}{0.012}                        \\\cline{1-2}\hline\hline
                         \multirow{4}{*}{Bartell Index 60}    & \multicolumn{1}{l|}{$\beta_0$}    & 2.349                      & 0.235                   & \multicolumn{1}{l|}{\textless.001}                & 1.721                      & 0.181                   & \multicolumn{1}{l|}{\textless.001}                & 2.436                      & 0.227                   & \multicolumn{1}{l|}{\textless.001}                & 1.701                      & 0.167                   & \multicolumn{1}{l|}{\textless.001}                & 1.829                      & 0.174                   & \multicolumn{1}{l|}{\textless.001}                & 1.841                      & 0.182                   & \multicolumn{1}{l|}{\textless.001}                & 2.010                      & 0.198                   & \multicolumn{1}{l|}{\textless.001}                \\
   \multicolumn{1}{|l|}{}                      & \multicolumn{1}{l|}{$\beta_1$}    & -0.200                     & 0.264                   & \multicolumn{1}{l|}{0.449}                        & -0.036                     & 0.216                   & \multicolumn{1}{l|}{0.869}                        & -0.402                     & 0.284                   & \multicolumn{1}{l|}{0.157}                        & -0.016                     & 0.209                   & \multicolumn{1}{l|}{0.939}                        & -0.134                     & 0.212                   & \multicolumn{1}{l|}{0.527}                        & -0.070                     & 0.228                   & \multicolumn{1}{l|}{0.759}                        & -0.058                     & 0.247                   & \multicolumn{1}{l|}{0.814}                        \\
     \multicolumn{1}{|l|}{}                    & \multicolumn{1}{l|}{$\beta_2$}    & -0.472                     & 0.318                   & \multicolumn{1}{l|}{0.138}                        & -0.360                     & 0.248                   & \multicolumn{1}{l|}{0.146}                        & -0.494                     & 0.298                   & \multicolumn{1}{l|}{0.098}                        & -0.392                     & 0.236                   & \multicolumn{1}{l|}{0.096}                        & -0.466                     & 0.242                   & \multicolumn{1}{l|}{0.054}                        & -0.431                     & 0.245                   & \multicolumn{1}{l|}{0.078}                        & -0.369                     & 0.266                   & \multicolumn{1}{l|}{0.166}                        \\
     \multicolumn{1}{|l|}{}                    & \multicolumn{1}{l|}{$\beta_3$}    & 0.663                      & 0.380                   &\multicolumn{1}{l|}{0.081}                        & 0.662                      & 0.311                   & \multicolumn{1}{l|}{0.033}                        & 0.703                      & 0.375                   & \multicolumn{1}{l|}{0.061}                        & 0.683                      & 0.301                   & \multicolumn{1}{l|}{0.023}                        & 0.784                      & 0.327                   & \multicolumn{1}{l|}{0.016}                        & 0.734                      & 0.323                   & \multicolumn{1}{l|}{0.023}                        & 0.620                      & 0.337                   & \multicolumn{1}{l|}{0.066}                        \\ \cline{1-2}\hline\hline
\multirow{4}{*}{Bartell Index 95}    & \multicolumn{1}{l|}{$\beta_0$}    & 0.085                      & 0.161                   & \multicolumn{1}{l|}{0.598}                        & -0.098                     & 0.149                   & \multicolumn{1}{l|}{0.512}                        & -0.122                     & 0.149                   & \multicolumn{1}{l|}{0.413}                        & -0.125                     & 0.149                   & \multicolumn{1}{l|}{0.400}                        & -0.076                     & 0.140                   & \multicolumn{1}{l|}{0.591}                        & -0.065                     & 0.145                   & \multicolumn{1}{l|}{0.652}                        & -0.031                     & 0.149                   & \multicolumn{1}{l|}{0.838}                        \\
   \multicolumn{1}{|l|}{}                      & \multicolumn{1}{l|}{$\beta_1$}    & 0.288                      & 0.161                   & \multicolumn{1}{l|}{0.073}                        & 0.258                      & 0.155                   & \multicolumn{1}{l|}{0.097}                        & 0.216                      & 0.157                   & \multicolumn{1}{l|}{0.170}                        & 0.282                      & 0.154                   &\multicolumn{1}{l|}{0.066}                        & 0.241                      & 0.151                   & \multicolumn{1}{l|}{0.110}                        & 0.312                      & 0.157                   & \multicolumn{1}{l|}{0.047}                        & 0.305                      & 0.152                   & \multicolumn{1}{l|}{0.044}                        \\
    \multicolumn{1}{|l|}{}                     & \multicolumn{1}{l|}{$\beta_2$}    & -0.163                     & 0.236                   & \multicolumn{1}{l|}{0.489}                        & -0.219                     & 0.214                   & \multicolumn{1}{l|}{0.304}                        & -0.197                     & 0.210                   & \multicolumn{1}{l|}{0.349}                        & -0.204                     & 0.217                   & \multicolumn{1}{l|}{0.346}                        & -0.211                     & 0.206                   & \multicolumn{1}{l|}{0.305}                        & -0.167                     & 0.213                   & \multicolumn{1}{l|}{0.434}                        & -0.144                     & 0.216                   & \multicolumn{1}{l|}{0.505}                        \\
    \multicolumn{1}{|l|}{}                     & \multicolumn{1}{l|}{$\beta_3$}    & 0.505                      & 0.242                   & \multicolumn{1}{l|}{0.037}                        & 0.558                      & 0.232                   & \multicolumn{1}{l|}{0.016}                        & 0.576                      & 0.228                   & \multicolumn{1}{l|}{0.012}                        & 0.519                      & 0.228                   & \multicolumn{1}{l|}{0.023}                        & 0.540                      & 0.230                   & \multicolumn{1}{l|}{0.019}                        & 0.526                      & 0.231                   & \multicolumn{1}{l|}{0.023}                        & 0.538                      & 0.230                   & \multicolumn{1}{l|}{0.020}                        \\ \cline{1-2}\hline\hline
\multirow{4}{*}{mental health score}    & \multicolumn{1}{l|}{$\beta_0$}    & 46.139                     & 0.810                   & \multicolumn{1}{l|}{\textless.001}                & 45.106                     & 0.926                   & \multicolumn{1}{l|}{\textless.001}                & 46.411                     & 1.206                   & \multicolumn{1}{l|}{\textless.001}                        & 44.656                     & 0.865                   & \multicolumn{1}{l|}{\textless.001}                & 45.683                     & 0.833                   & \multicolumn{1}{l|}{\textless.001}                & 45.174                     & 0.830                   & \multicolumn{1}{l|}{\textless.001}                & 45.146                     & 0.813                   & \multicolumn{1}{l|}{\textless.001}                \\
\multicolumn{1}{|l|}{}                         & \multicolumn{1}{l|}{$\beta_1$}    & 3.320                      & 0.931                   & \multicolumn{1}{l|}{\textless.001}                & 3.426                      & 1.074                   & \multicolumn{1}{l|}{0.001}                        & 2.476                      & 1.222                   & \multicolumn{1}{l|}{0.043}                        & 3.826                      & 0.983                   & \multicolumn{1}{l|}{\textless.001}                & 2.656                      & 0.963                   & \multicolumn{1}{l|}{0.006}                        & 3.544                      & 0.977                   & \multicolumn{1}{l|}{\textless.001}                & 3.259                      & 0.943                   & \multicolumn{1}{l|}{0.001}                        \\
 \multicolumn{1}{|l|}{}                        & \multicolumn{1}{l|}{$\beta_2$}    & 0.017                      & 1.201                   & \multicolumn{1}{l|}{0.989}                       & -0.404                     & 1.236                   & \multicolumn{1}{l|}{0.744}                        & 0.002                      & 1.118                   & \multicolumn{1}{l|}{0.998}                        & -0.117                     & 1.212                   & \multicolumn{1}{l|}{0.923}                        & -0.509                     & 1.226                   & \multicolumn{1}{l|}{0.678}                        & -0.119                     & 1.255                   & \multicolumn{1}{l|}{0.924}                        & -0.235                     & 1.219                   & \multicolumn{1}{l|}{0.847}                        \\
 \multicolumn{1}{|l|}{}                        & \multicolumn{1}{l|}{$\beta_3$}    & -0.067                      & 1.383                   & \multicolumn{1}{l|}{0.962}                        & 0.671                      & 1.435                   & \multicolumn{1}{l|}{0.640}                        & -0.092                     & 1.386                   & \multicolumn{1}{l|}{0.947}                        & 0.553                      & 1.385                   & \multicolumn{1}{l|}{0.690}                        & 0.954                      & 1.403                   & \multicolumn{1}{l|}{0.497}                        & 0.303                      & 1.426                   & \multicolumn{1}{l|}{0.832}                        & 0.865                      & 1.339                   & \multicolumn{1}{l|}{0.518}                        \\ \cline{1-2}\hline\hline
\multirow{4}{*}{physical health score}    & \multicolumn{1}{l|}{$\beta_0$}    & 46.573                     & 0.840                   & \multicolumn{1}{l|}{\textless.001}                & 45.145                     & 0.806                   & \multicolumn{1}{l|}{\textless.001}                & 46.128                     & 0.951                   & \multicolumn{1}{l|}{\textless.001}                & 45.035                     & 0.805                   & \multicolumn{1}{l|}{\textless.001}                & 45.258                     & 0.786                   & \multicolumn{1}{l|}{\textless.001}                & 45.383                     & 0.773                   & \multicolumn{1}{l|}{\textless.001}                & 45.315                     & 0.802                   & \multicolumn{1}{l|}{\textless.001}                \\
   \multicolumn{1}{|l|}{}                          & \multicolumn{1}{l|}{$\beta_1$}    & -3.928                     & 0.808                   & \multicolumn{1}{l|}{\textless.001}                & -3.310                     & 0.846                   & \multicolumn{1}{l|}{\textless.001}                & -3.918                     & 0.892                   & \multicolumn{1}{l|}{\textless.001}                & -3.267                     & 0.833                   & \multicolumn{1}{l|}{\textless.001}                & -3.258                     & 0.836                   & \multicolumn{1}{l|}{\textless.001}                & -3.301                     & 0.803                   & \multicolumn{1}{l|}{\textless.001}                & -3.275                     & 0.819                   & \multicolumn{1}{l|}{\textless.001}                \\
     \multicolumn{1}{|l|}{}                        & \multicolumn{1}{l|}{$\beta_2$}    & -0.527                     & 1.226                   & \multicolumn{1}{l|}{0.667 }                        & -0.481                     & 1.169                   & \multicolumn{1}{l|}{0.681}                        & -0.284                     & 1.128                   & \multicolumn{1}{l|}{0.801}                        & -0.518                     & 1.165                   &\multicolumn{1}{l|}{0.656}                        & -0.863                     & 1.131                   & \multicolumn{1}{l|}{0.446}                        & -0.645                     & 1.159                   & \multicolumn{1}{l|}{0.578}                        & -0.385                     & 1.155                   & \multicolumn{1}{l|}{0.739}                        \\
    \multicolumn{1}{|l|}{}                         & \multicolumn{1}{l|}{$\beta_3$}    & 3.035                     & 1.206                   & \multicolumn{1}{l|}{0.012}                        & 3.148                      & 1.244                   & \multicolumn{1}{l|}{0.011}                        & 2.944                      & 1.237                   & \multicolumn{1}{l|}{0.017}                        & 3.343                      & 1.232                   & \multicolumn{1}{l|}{0.007}                        & 3.505                      & 1.210                   & \multicolumn{1}{l|}{0.004}                        & 3.340                      & 1.248                   & \multicolumn{1}{l|}{0.007}                        & 3.015                      & 1.193                   & \multicolumn{1}{l|}{0.012}                        \\ \cline{1-2}\hline
\end{tabular}}
\captionof{table}{A comparison of point estimates of coefficients, standard deviations and p-values of the five models of interest under seven treatments of missing data in the original QASC data set.}
\label{table_exp_settings}
}}

\section{Discussion}
In this paper, we developed a copula based imputation model for multilevel data sets with mixed data. Copula based imputation models have a sound theoretical foundation and we have shown through simulations that copula based imputation models achieve reasonably accurate predictions of the missing values and recovery of parameters in some models of interest.

The copula based imputation models outperform the competing methods, especially when the variable distributions depart from normality. We also recommend taking into account clustering effects to incorporate information from the grouping structure in the analysis. This is confirmed from our simulation results, that when the ICC is high, imputation models with random effects added achieve better results.

One extension to our models is to add some `fixed' covariates. For the copula models in Section 3, all the variables appear on one side of the equations in (\ref{combined}) and we model their relationship through the correlation matrices on the latent variable scale. But it is often of interest to see both the relationship among variables on the response side and the relationship between the responses and some covariates. For example in the QASC data set, `treatment' is fixed by design at hospital level, so we can treat it as a regressor. By doing so, the treatment effects on some process of care variables can be detected directly through the copula model on the latent variable scale. Here we consider variables with ordering, and extension to nominal variables is straightforward. Let $i={1,...,m}$ be the group index, $j={1,...,n_i}$ be the individual index within group $i$, and $l={1,...,p}$ be the variable index. Suppose the first $k$ variables have common covariates $x_{i1},...,x_{iq}$ at the group level, in other words, they are fixed within group $i$. The correlation matrices for residual and random effects $b_i$ are $\Gamma$ and $\Psi$ respectively as before, but the mean of the latent variables $z$ is no longer zero. Again we use the monotone transformation $z_{ijl}=\Phi^{-1}(F(y_{ijl}))$ to obtain the extended rank likelihood, then the model becomes:

\begin{equation}\label{hierarchical copula}
\begin{aligned}
  &z_{ij}\sim N(b_i+x_i(\beta, 0), \Gamma), b_i\sim N(0,\Psi)\\
  &\Updownarrow\\
  &(z_{ij1},...,z_{ijk},...,z_{ijp})\sim N\Big(b_i+(x_{i1},...,x_{iq})\begin{pmatrix}
  \beta_{11} &  \cdots & \beta_{1k} &0&\cdots&0\\
  \vdots    &  & \vdots & \vdots&& \vdots \\
  \beta_{q1} & \cdots & \beta_{qk} &0&\cdots&0
 \end{pmatrix},\Gamma)\Big)
\end{aligned}
\end{equation}
It is straightforward to derive the full conditional distributions for the Gibbs sampler, and we omit the details here.

Choosing the form of copula is another issue which is a critical yet complicated task. \cite{kole2007selecting}, \cite{trivedi2007copula} provide some guidance on choosing among existing copulas or creating new families of copulas. In this paper, we focused on the Gaussian copula because it is easy to extend to higher dimensions and computationally convenient. However, the main drawbacks of the Gaussian copula are the symmetry assumption and absence of tail dependence \citep{demarta2005t}. Therefore, some goodness-of-fit tests should be examined to check for a need to use other forms of copulas, for example, a (mixture of skewed) t-copulas.

\bibliographystyle{kbib}

\end{document}